\documentclass[journal]{IEEEtran}
\usepackage{subfigure}
\usepackage{graphicx}
\usepackage{epsfig,latexsym,amsmath,graphics}
\usepackage{amssymb,dsfont,amsthm}
\usepackage{cite,color}
\usepackage{url}
\usepackage{booktabs}
\usepackage{algorithm, algorithmicx}
\usepackage{algpseudocode}
\usepackage{stfloats}
\usepackage{verbatim}
\usepackage{bm}
\usepackage{stfloats}
\usepackage{multirow}
\usepackage{ulem}
\usepackage{caption}
\usepackage{booktabs} 
\usepackage{stfloats}
\usepackage{float}
\usepackage{cases}
\usepackage{diagbox} % 加载宏包

\begin{document}
\pdfoutput=1
\title{Characterization of Rydberg-Atom Signal Reception of Dual-Frequency Signals Coupled with Two Energy Levels}

%\author{Hao~Wu,% <-this % stops a space
	%
	%}

\author{Hao Wu, Chongwu Xie, Xinyuan Yao, Kang-Da Wu, Shanchi Wu, Rui Ni, Guo-Yong Xiang, Chen Gong
	% <-this % stops a space
	\thanks{This work was supported in part by National Natural Science Foundation of China under Grant 62171428.}
	\thanks{Hao Wu, Xinyuan Yao, Shanchi Wu, Chen Gong are with the School of Information Science and Technology in University of Science and Technology of China, Email address: \{wuhao0719, yxy200127, wsc0807\}@mail.ustc.edu.cn,cgong821@ustc.edu.cn.
		
	Chongwu Xie, Kang-Da Wu and Guo-Yong Xiang are with the CAS Key Laboratory of Quantum Information in University of Science and Technology of China, and Guo-Yong Xiang is with Hefei National Laboratory, Email address: cwxie@mail.ustc.edu.cn, \{wukangda, gyxiang\}@ustc.edu.cn.
		
		Rui Ni is with Huawei Technology, Email address: raney.nirui@huawei.com.}% <-this % stops a space
}

\maketitle

\begin{abstract}
Rydberg atomic sensors have been adopted for novel radio frequency (RF) measurement technique and the sensing capability for signals in multiple frequencies makes it attractive for multi-user communication. However, unlike traditional antennas where the signals in multiple frequencies are orthogonal, the received signals of atomic sensors corresponding to different energy levels will be downconverted to the baseband simultaneously, resulting in multi-user interference. Thus, in this paper, we analyze the mutual interference characteristics of two RF signals with different carrier frequencies coupling different energy levels. We introduce the joint response coefficient based on the receiver characteristics and analyze the interference of one user to another. We analyze the bit-error rate (BER) and symbol-error rate (SER) for two signals coupling two different energy levels. We also conduct experiments to validate the BER and SER results.
\end{abstract}


\maketitle

\section{Introduction}
Rydberg atoms have been adopted for electric field measurement, which can lead to a direct International System of Units (SI) traceable and self-calibrated measurement paradigm. Rydberg atoms, characterized by highly excited electronic states with large principal quantum numbers, exhibit exceptional electric field sensitivity. This unique property makes them ideal candidates for developing atomic sensors capable of detecting and demodulating communication signals. Rydberg atom-based electric field measurement techniques have achieved remarkable progress, with the measurement sensitivity currently reaching nV/cm/$\sqrt{\text{Hz}}$ \cite{anderson2016optical,paradis2019atomic,jing2020atomic,gordon2019weak,borowka2024continuous,knarr2023spatiotemporal}. Furthermore, these techniques demonstrate exceptional frequency bandwidth coverage, extending from sub-kilohertz (kHz) regimes to gigahertz (GHz) and terahertz (THz) domains\cite{jau2020vapor,li2023super,downes2020full,wade2017real}. It can also measure the polarization of radio frequency (RF) electric field\cite{sedlacek2013atom,yuan2024isotropic,yin2024measurement,wang2023precise}. At present, the recognition of multi-frequency microwave electric field has been realized\cite{liu2022deep,zhang2022rydberg}. These have opened up other types of sensing, including direction of arrival of an radio frequency field\cite{robinson2021determining}, subwavelength imaging\cite{holloway2014sub,holloway2017atom,downes2020full} and near-field antenna pattern measurements\cite{simons2019embedding,shi2023near}.

The signal reception of Rydberg atomic sensors is based on discrete energy levels. Thus, it is possible to simultaneously receive multiple carrier signals with large frequency span by coupling different atomic energy levels. For example, in work \cite{meyer2023simultaneous}, the authors demonstrate simultaneous demodulation and detection of five RF signals spanning nearly two decades (6 octaves), from 1.7 GHz to 116 GHz. Other basic researches related to the atomic multi-levels have been reported in works\cite{robinson2021atomic,you2023exclusive,rotunno2023investigating,berweger2023rydberg,allinson2024simultaneous}. In the field of communication\cite{wen2024rydberg}, based on multi-level receiving characteristics, a frequency-hopping Rydberg receiver has been realized. The digital transmission with a data rate of 1 Mbps is performed reliably, within a tunable bandwidth of 50 MHz and a max hopping rate of 20000 hop/s. Thus, the atomic sensors, with abundant discrete receiving bandwidth windows, show great potential for applications in multi-frequency reception.

However, unlike traditional dipole antennas, the received signal corresponding to different energy levels will be downconverted to the baseband simultaneously, resulting in the mutual infection among multiple users, even under different wavelengths. In work\cite{wu2023linear}, the nonlinear characteristics and gain performance as two fixed RF signals (local oscillator and modulated signal) coupled to the same atomic energy level is investigated. The response of random signal and multi-levels would be complex for atomic sensors. In addition, the diverse atomic energy level structures formed by multiple subcarriers with different frequencies make nonlinear analysis more complicated.

In this work, we investigate the mutual interference between two users with two carrier frequencies coupling two energy levels. We model and simplify the atomic sensors as an envelope receiver based on density matrix and master equation. We analyze the bit error rate (BER) and symbol error rate (SER) under interference cases. We also conduct experiments to validate the BER results from theoretical analysis.

The remainder of this paper is organized as follows. In Section \ref{T2}, we introduce the multiple frequency model and the mutual interference theoretically. In Section \ref{T3}, we analyze the BER and SER performance with synchronous On-Off Keying (OOK) signals. In Section \ref{T4}, the experimental results are shown and compared with theoretical results. Finally, we conclude this paper in Section \ref{T5}.

\section{Signal and Receiving Model}\label{T2}
\subsection{Signal Model}\label{T21}

\begin{figure*}
	\centering
	\includegraphics[width=0.95\textwidth]{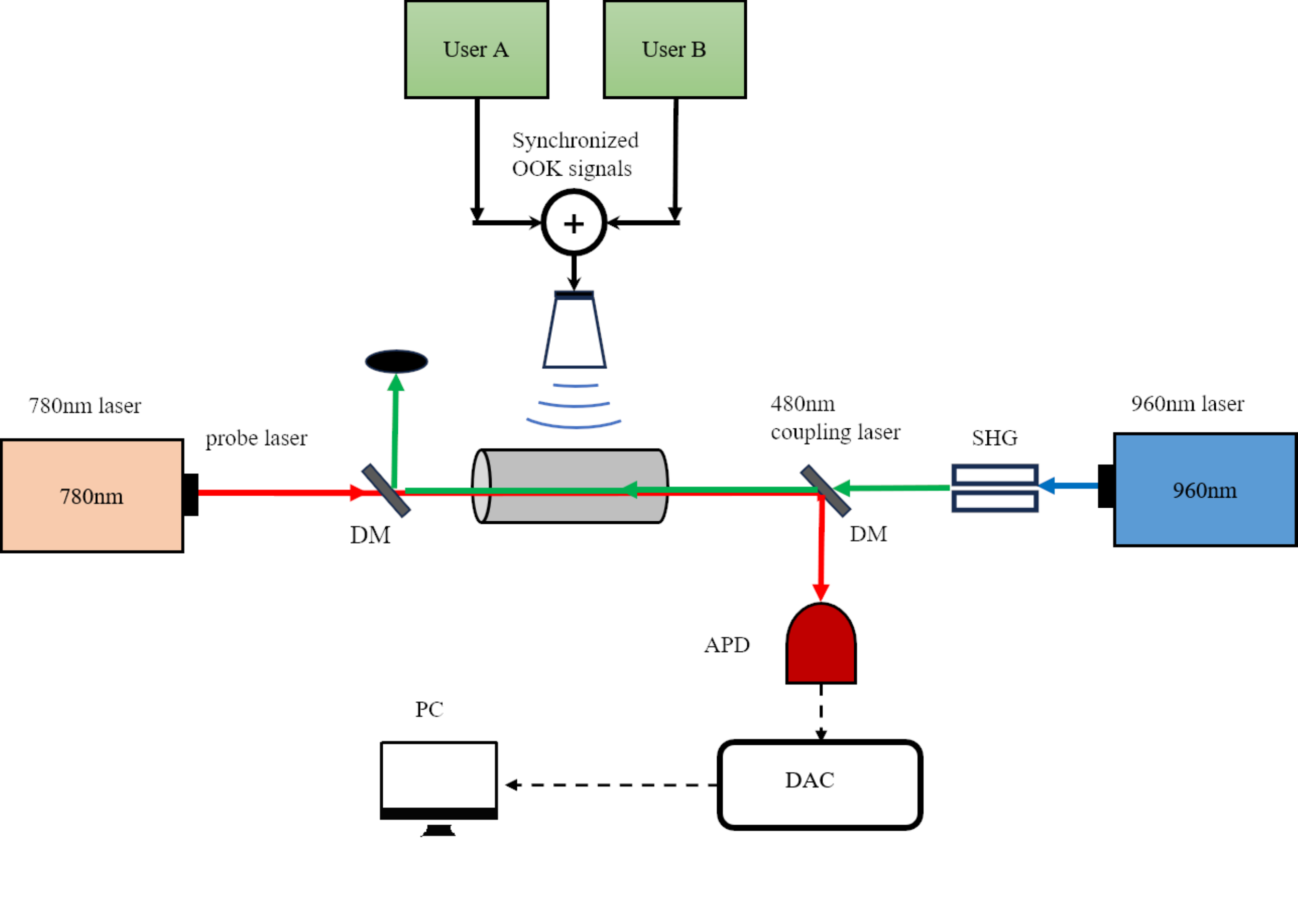}
	\caption{The diagram of multi-user Rydberg atomic detection system.}
	\label{systemdiagram}
\end{figure*}

$\ \ $ A schematic diagram of the detection system based on Rydberg atomic sensor with multi-user input is shown in Fig. \ref{systemdiagram}. We employ a 780 nm laser as the probe laser and a 480 nm laser as the coupling laser. These lasers propagate in opposite directions and excite part of atoms inside the atomic cell to Rydberg states. Eventually, the probe and coupling laser are separated by two dichroic mirrors (DM). The output probe laser, passing through atomic cell, is received by an avalanche photodiode (APD). In the experiment, the two users A and B generate the synchronized OOK signals, and a microwave combiner mix and send the signals from two users. The data is collected by a data acquisition card (DAC) and stored in personal computer (PC) for offline symbol detection. In this work, we assume that the probe laser, coupling laser, and two RF fields share the same polarization. 

In order to simplify the model, we assume the two users send synchronous OOK signal $E_{s1}\left( t \right)$ and $E_{s2}\left( t \right)$, respectively, given by
\begin{equation}
	\begin{aligned}
		E_{s1}\left( t \right)=
		\begin{cases}
			\left|E_{s1}\right|\sin \left( 2\pi f_{s1}t+\phi _{s1} \right) & \text{bit} = 1   \\
			0 & \text{bit} = 0 
		\end{cases},
	\end{aligned}
\end{equation}	
\begin{equation}
	\begin{aligned}
		E_{s2}\left( t \right)=
		\begin{cases}
			\left|E_{s2}\right|\sin \left( 2\pi f_{s2}t+\phi _{s2} \right) & \text{bit} = 1     \\
			0 & \text{bit} = 0  
		\end{cases},
	\end{aligned}
\end{equation}	
where the amplitude of two OOK signals with carrier frequencies $f_{s1}$ and $f_{s2}$ are $\left|E_{s1}\right|$ and $\left|E_{s2}\right|$ as the bits are one, respectively. Let $\phi_{s1}$ and $\phi_{s2}$ denote the corresponding phases. Without loss of generality, we set $\phi_{s1}=\phi_{s2}=0$.

%\begin{figure}[htbp]
%	\centering
%	\includegraphics[width=0.42\textwidth]{fig2//PYAG.pdf}
%	\caption{Antenna gain with respect to signal frequency.}
%	\label{AG}
%\end{figure}

%Figure. \ref{AG} shows the difference between traditional dipole antenna and Rydberg atomic sensors and a part of acceptable frequencies of Rb atoms $53D_{5/2}$. 
%For traditional dipole antennas, the frequencies of two RF signals should be close to the resonance frequency of dipole antenna. But, another signal with different frequency can also affect the atomic sensors by coupling other atomic energy levels.
Based on the discrete frequency window of Rydberg atoms, the atomic sensors have the capability of perceiving frequencies $f_1 = 14.23$ GHz and $f_2 = 15.59$ GHz. In the following part, we analyze the interference characteristics between these windows. Specifically, one of the user sends OOK signal with frequency $f_1 = 14.23$ GHz coupling transition $53D_{5/2}-54P_{3/2}$ and another one sends with frequency $f_2 = 15.59$ GHz coupling transition $53D_{5/2}-52F_{7/2}$.  

\subsection{Atomic Energy Level Model}\label{T22}

\begin{figure}[htbp]
	\centering  %图片全局居中
	\label{DoubleModelCase2}
	\includegraphics[width=0.45\textwidth]{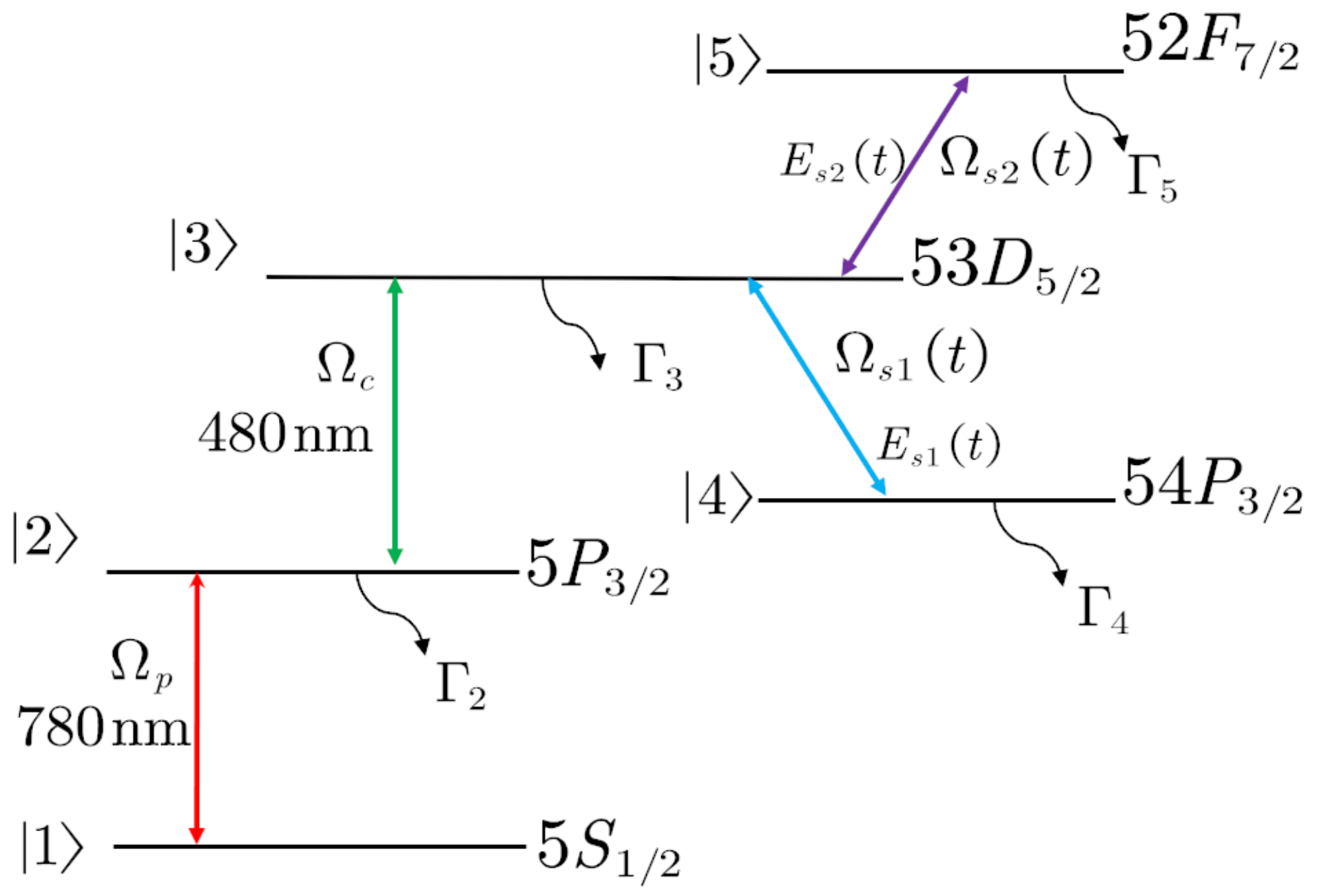}
	\caption{The structure of five energy level model.}
	\label{DoubleModel}
\end{figure}

$\ \ $ Here, we consider a five energy level model, due to two users coupling different atomic energy levels, Hamiltonian ${H}$ and Lindblad operator ${\mathcal{L}}$ are both characterized as  $5\times5$ matrices. The typical energy level structure is shown in Fig. \ref{DoubleModel}. The probe and coupling laser beams excite the atoms from state $\left| 1 \right> $ to state $\left| 2 \right> $ and from state $\left| 2 \right> $ to state $\left| 3 \right>$, respectively. The two signals $E_{s1}$ and $E_{s2}$ couple the Rydberg transition from state $\left| 3 \right>$ to state $\left| 4 \right> $ and from state $\left| 3 \right>$ to state $\left| 5 \right> $, respectively. The detailed model description is provided in the Appendix\ref{A}.

%we denote $G[\left| E_{s2} \right| ,  \left| E_{s1} \right|]$ as the joint response coefficient to show the ouput APD voltage of the atomic system given $\left| E_{s1} \right|$ and $\left| E_{s2} \right|$. W

%let $V[\left| E_{s1} \right| ,  \left| E_{s2} \right|]$ denote the APD output voltage corresponding to the given input field $\left| E_{s1} \right|$ and $\left| E_{s2} \right|$.we denote $V[\left| E_{s1} \right| ,  \left| E_{s2} \right|]$ as the joint response coefficient to show the ouput APD voltage of the atomic system given $\left| E_{s1} \right|$ and $\left| E_{s2} \right|$

For two users with signal electric field intensity $\left| E_{s1} \right|$ and $\left| E_{s2} \right|$, we define $V[\left| E_{s1} \right| ,  \left| E_{s2} \right|]$ as the APD output voltage given input field $\left| E_{s1} \right|$ and $\left| E_{s2} \right|$. Without loss of generality, we divide the APD voltage by the difference between two extreme responses to obtain the normalized joint response coefficient $G\left[ \left| E_{s1} \right|,\left| E_{s2} \right| \right]$, which is given by
\begin{equation}
	\begin{aligned}
		G\left[ \left| E_{s1} \right|,\left| E_{s2} \right| \right] =\frac{V\left[ \left| E_{s1} \right|,\left| E_{s2} \right| \right] -V_s}{V_0-V_s},
	\end{aligned}
\end{equation}
where $V_0=V[0,0]$ is the APD voltage without electric field and $V_s=\lim_{\left| E \right|\rightarrow +\infty}V\left[ \left| E \right|,\cdot \right]=\lim_{\left| E \right|\rightarrow +\infty}V\left[ \cdot,\left| E \right| \right]$ is the the APD saturation output under large electric field. Due to normalization being a scaling of signal strength, including noise, this operation does not affect the results. For the normalized joint response coefficient $G$, we have  
\begin{equation}\label{Boundary}
	\begin{aligned}
		G\left[ 0,0 \right] &=1,
		\\
		\lim_{|E_{s1}|\rightarrow +\infty}G\left[ |E_{s1}|,\cdot \right] &=0,
		\\
		\lim_{|E_{s2}|\rightarrow +\infty}G\left[\cdot,|E|_{s2} \right] &=0.
	\end{aligned}
\end{equation}

Based on the typical five level Rydberg atomic model in the Appendix\ref{A}, in Fig. \ref{modelshow2}, we simulate the relationship between intensity $|E_{s2}|$ and normalize APD voltage to illustrate the mutual interference under different intensity $|E_{s1}|$. The parameters are the same as the experiment in the paper. We consider the response coefficient $G[\left| E_{s1} \right| ,  \left| E_{s2} \right|]$ under three electric field intensity $\left| E_{1} \right|<\left| E_{2} \right|<\left| E_{3} \right|$, and the output normalized APD voltage $V_{APD}(t)$ is blocked with the increasing intensity $|E_{s1}|$, which flatten the APD voltage curve to reduce the effective APD output range and leads a lower signal gain. Thus, it can be foreseen that when one of the two users has higher signal power, the communication of another one will be affected.

\begin{figure}[htbp]
	\centering  %图片全局居中
	\includegraphics[width=0.4\textwidth]{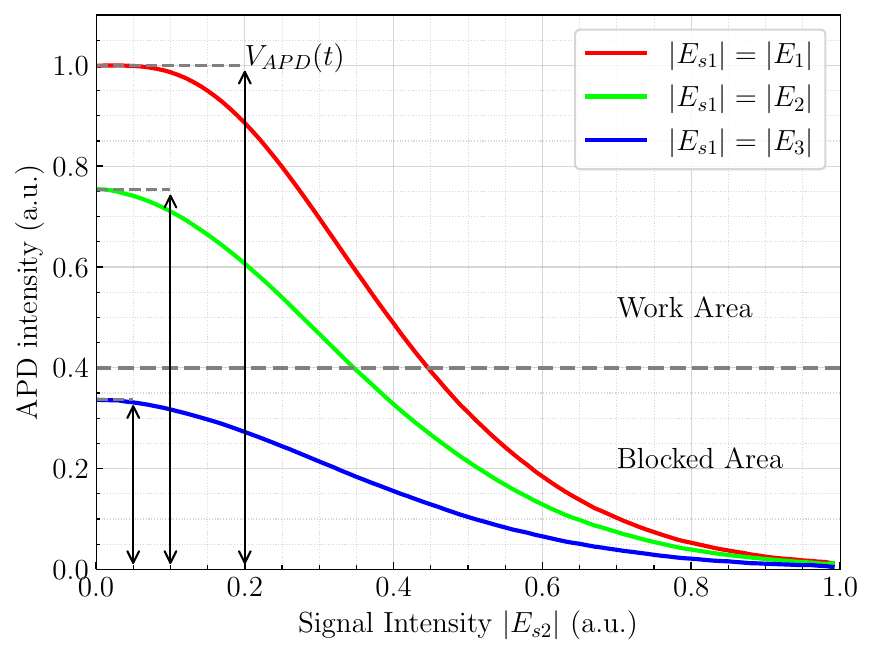}
	\caption{APD voltage with respect to electric field intensity $\left| E_{s2} \right|$ given $\left| E_{s1} \right|$. Red: $G[\left| E_{s1} \right|=\left| E_{1} \right|,\left| E_{s2} \right| ]$. Green: $G[\left| E_{s1} \right|=\left| E_{2} \right|,\left| E_{s2} \right| ]$. Blue: $G[  \left| E_{s1} \right|=\left| E_{3} \right|,\left| E_{s2} \right|]$.}
	\label{modelshow2}
\end{figure}

\section{BER Performance of Two Users}\label{T3}

\begin{table*}
	\caption{The bits probability and combinations under different cases and symbols\label{CSPC}}
	\centering
	\begin{tabular}{|c|c|c|c|c|c|c|}
		\hline
		\diagbox{Cases}{Bits-Probability}{Combinations} & $G\left[ |E_{s1}|,|E_{s2}| \right] $ & $G\left[ |E_{s1}|,0 \right] $ & $G\left[ 0,|E_{s2}| \right] $ & $G\left[ 0,0 \right] $ \\
		\hline
		Case 1 & One-0.5 & Zero-0.5 & - & - \\
		\hline
		Case 2 & One-0.5 & - & Zero-0.5&- \\
		\hline
		Case 3 &One-0.25  & Zero-0.25 & One-0.25 & Zero-0.25\\
		\hline
		Case 4 & One-0.25 & One-0.25  & Zero-0.25&	Zero-0.25\\
		\hline
	\end{tabular}
\end{table*}

$\ \ $ In this section, we consider the users with two synchronized OOK signals $E_{s1}$ and $E_{s2}$ ($|E_{s1}|>|E_{s2}|$) and their carrier frequencies $f_{s1}$ and $f_{s2}$. We adopt the maximum likelihood detection and analyze the approximate BER performance with Gaussian noise under four cases, as well as the SER of the two users. 

%For CELM, we ignore the impact of carrier frequency on normalized gain coefficient $G[\cdot]$, i.e., the frequencies $f_{s1}$ and $f_{s2}=f_{s1}+\Delta f$ have the same $G[\cdot]$. Because the ultimate APD voltage $P_{APD}(t)$ is the sum of DC and AC parts, we adopt fourier transform to extract the intensity of the two parts. Denote $\sigma _{DC}^2$ and $\sigma _{AC}^2$ are independent the Gaussian noise variance of DC part and AC part, respectively. The BER can be calculated by the two dimensional four symbols detection model with mean $\left[ G\left[ 0\right],0 \right] $, $\left[ G\left[ |E_{s1}|\right],0 \right] $, $\left[G\left[ |E_{s2}|\right],0 \right] $, and $\left[ G\left[ |E_{s1}|\right],G'\left[ |E_{s1}| \right] |E_{s2}| \right] $.

\begin{figure}[htbp]
	\centering  %图片全局居中
	\includegraphics[width=0.4\textwidth]{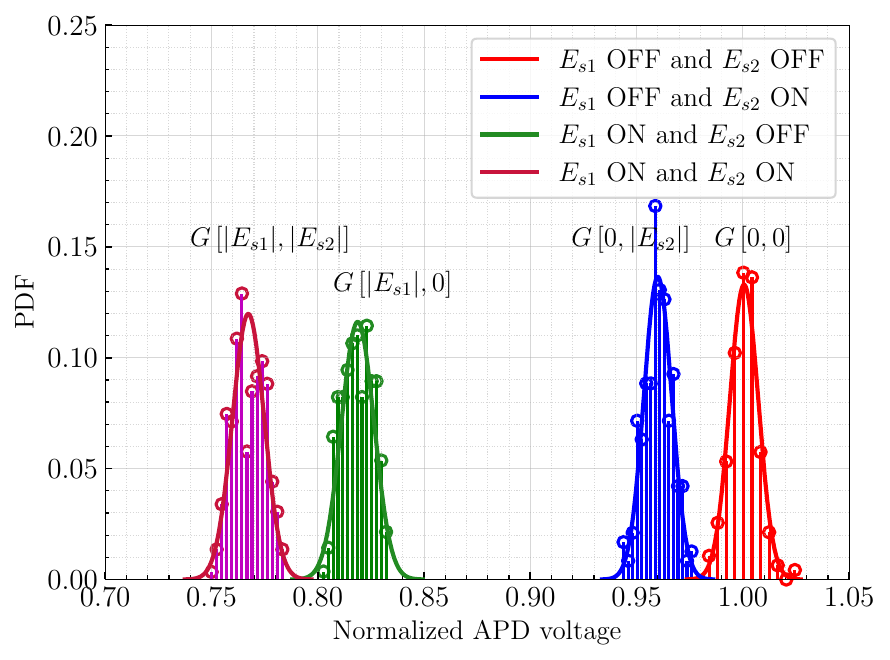}
	\caption{The statistical distribution of four APD voltage combinations in actual system. We set $E_{s1}=1.3$ mV/cm, $E_{s2}=1.3$ mV/cm, $f_{s1}=f_1$, and $f_{s2}=f_2$.}
	\label{IQI14S15}
\end{figure}

Specifically, we denote $G\left[ |E_{s1}|,|E_{s2}| \right] $ as the normalized joint response coefficient of the sensors when the two E-filed driving the Rydberg transition from state $53D_{5/2}$ to $54P_{3/2}$ and from state $53D_{5/2}$ to $52F_{7/2}$ are $|E_{s1}|$ and $|E_{s2}|$, respectively. Because two users send OOK signals coupling two atomic energy levels, the direct-current (DC) component of APD voltage can be used for symbol detection. After the atomic sensors output stabilizes, there are four combinations of APD average voltage: $G\left[ 0,0 \right] $, $G\left[ |E_{s1}|,0 \right] $, $G\left[ 0,|E_{s2}| \right] $, and $G\left[ |E_{s1}|,|E_{s2}| \right] $, where $G\left[ 0,0 \right] $ and $G\left[ |E_{s1}|,|E_{s2}| \right] $ represent two users sending bits zero and one, respectively. $G\left[ |E_{s1}|,0 \right] $ and $G\left[ 0,|E_{s2}| \right] $ represent that one user sending bit one and the other sending zero. In Fig. \ref{IQI14S15}, we test the statistical distribution of the four normalized voltage combinations in actual systems. Due to the presence of noise in the system, each combination exhibits a Gaussian distribution $\mathcal{N}\left(\mu_{\text{comb}},\sigma \right) $ with mean $\mu_{\text{comb}}$ and similar variances $\sigma$ approximatively.
%Under Gaussian noise, the BER can be calculated by one dimensional four symbols detection model with variance $\sigma ^2$ and mean $G\left[ 0,0 \right] $, $G\left[ |E_{1}|,0 \right] $, $G\left[ 0,|E_{2}| \right] $, and $G\left[ |E_{1}|,|E_{2}| \right] $.

\begin{figure*}
	\centering
	\subfigure[]{
		\begin{minipage}[t]{0.5\linewidth}
			\centering
			\includegraphics[width=3 in]{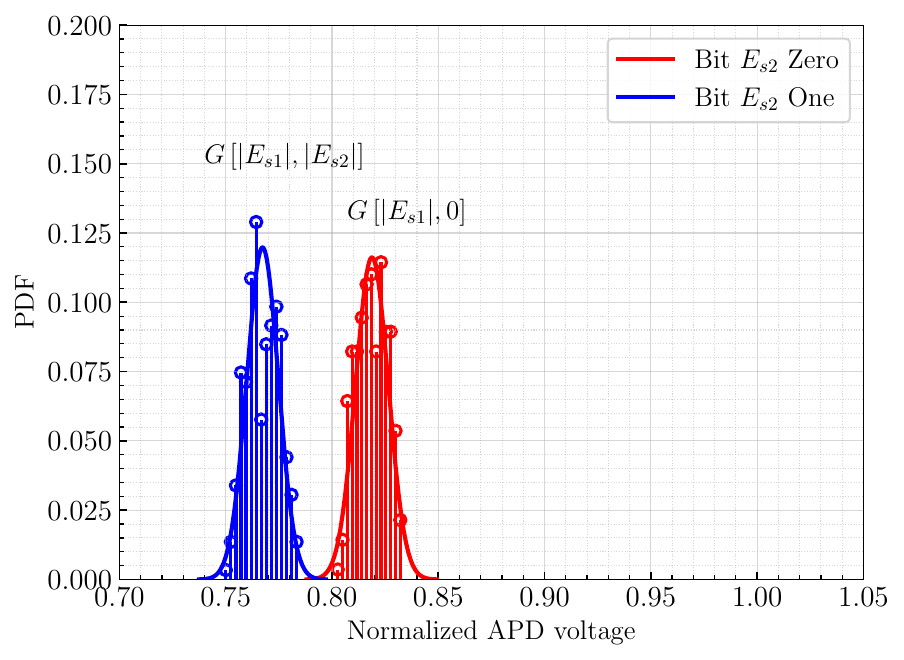}
			%\caption{Symbol switching from zero to zero}
			\label{PYDisCase1I14S15}
		\end{minipage}%
	}%
	\subfigure[]{
		\begin{minipage}[t]{0.5\linewidth}
			\centering
			\includegraphics[width=3 in]{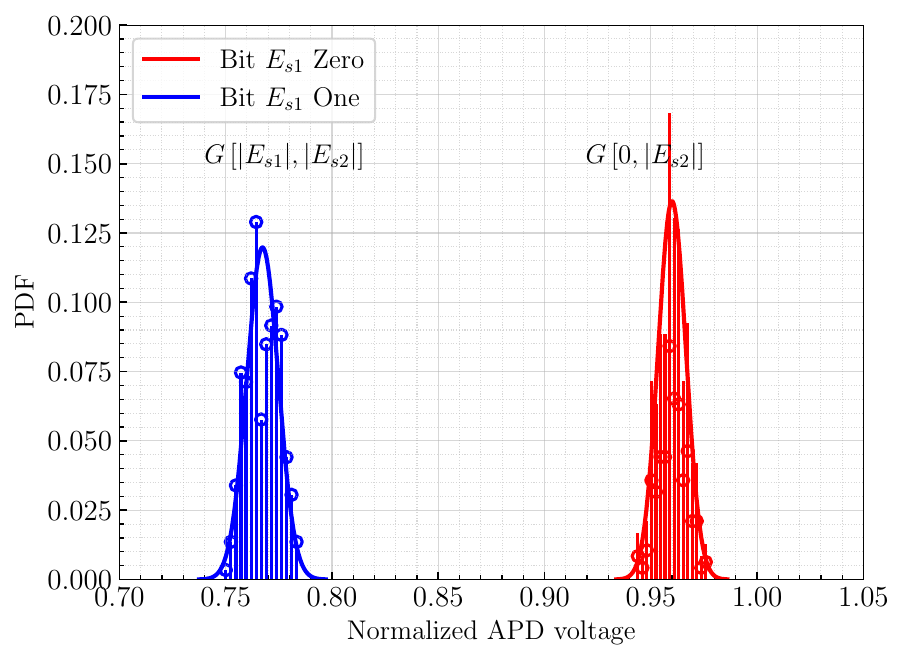}
			\label{PYDisCase2I14S15}
			%\caption{Symbol switching from zero to one}
		\end{minipage}%
	}%
	
	\subfigure[]{
		\begin{minipage}[t]{0.5\linewidth}
			\centering
			\includegraphics[width=3in]{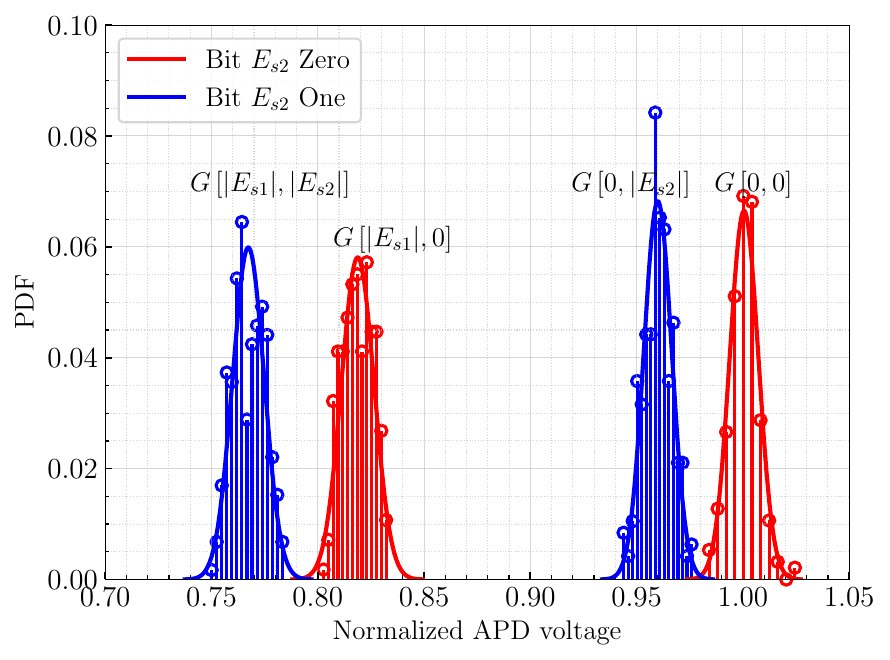}
			\label{PYDisCase3I14S15}
			%\caption{Symbol switching from one to zero}
		\end{minipage}
	}%
	\subfigure[]{
		\begin{minipage}[t]{0.5\linewidth}
			\centering
			\includegraphics[width=3in]{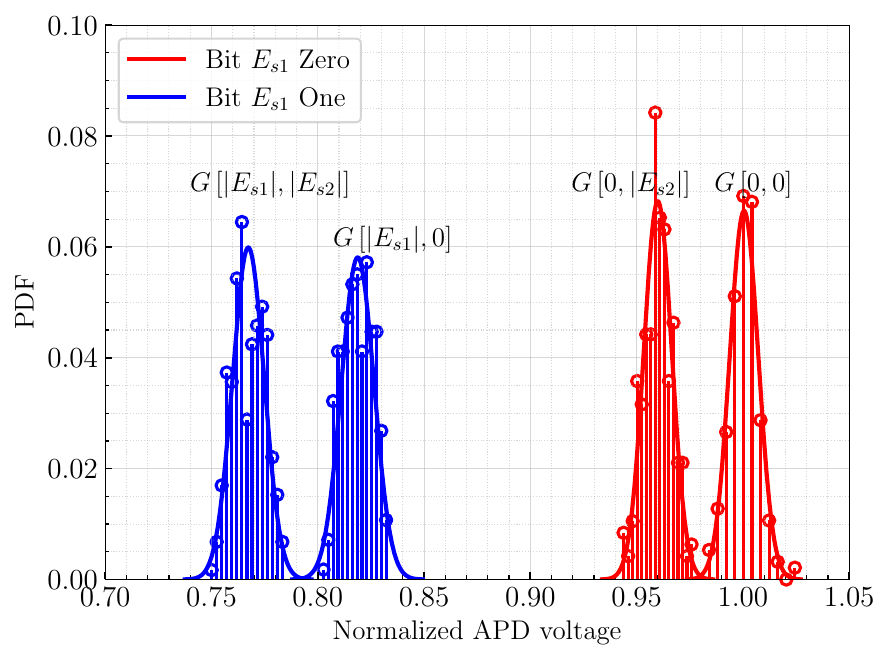}
			\label{PYDisCase4I14S15}
			%\caption{Symbol switching from one to one}
		\end{minipage}
	}%
	\centering
	\caption{The statistical distribution of combinations under four cases. The red and blue distributions represent bits zero and one, respectively. (a) Case 1. (b) Case 2. (c) Case 3. (d) Case 4.}
	\label{Dis}
\end{figure*}

Here, we consider the BER of one of the users under four cases corresponding to four types of two users communication scenarios, when another user sends a constant strength electric field without signal (Case 1 and Case 2) or a random OOK signal (Case 3 and Case 4). In Case 1, we consider the BER of weaker power random signal $E_{s2}$ under stronger constant intensity $\left| E_{s1}\right| $. In Case 2, we consider the BER of higher power random signal $E_{s1}$ under weaker constant intensity $\left| E_{s2}\right|$. In Case 3, we consider the BER of random signal $E_{s2}$ under random OOK signal $E_{s1}$. In Case 4, we consider the BER of random signal $E_{s1}$ under random OOK signal $E_{s2}$. The difference between Cases 1,2 and Cases 3,4 is that the former considers the interference of fixed strength electric field, while the latter considers the influence of random signals. In Fig. \ref{Dis} and Table \ref{CSPC}, we display the statistical distribution of bits zero and one under four cases and the figures help explain the calculation of BER.

\subsection{Case 1: The BER of Signal $E_{s2}$ with Constant Intensity $E_{s1}$.} \label{T31}

%
%Under CELM and a fixed E-field $|E_{s1}|$, the symbol one of $E_{s2}$ would form a difference frequency component in APD voltage but the symbol zero will not. Thus, Symbol zero and one of OOK signal $E_{s2}$ follow the Gaussian distribution with mean 0 and $G'\left[ |E_{s1}| \right] |E_{s2}|$, respectively. Under Gaussian noise, the BER can be expressed as
%
%\begin{equation}\label{C1BERCalculationFixSaF}
%	\begin{aligned}
	%		P_{C1,CELM}&=\frac{1}{2}\int_{th}^{+\infty}{\frac{1}{\sqrt{2\pi}\sigma _{AC}}e^{-\frac{\left( x-0 \right) ^2}{2\sigma _{AC}^{2}}}dx}
	%		\\
	%		&\ \ \ +\frac{1}{2}\int_{-\infty}^{th}{\frac{1}{\sqrt{2\pi}\sigma _{AC}}e^{-\frac{\left( x-G'\left[ |E_{s1}| \right] |E_{s2}| \right) ^2}{2\sigma _{AC}^{2}}}dx}
	%		\\
	%		&=Q\left( \frac{G'\left[ |E_{s1}| \right]}{2\sigma _{AC}}|E_{s2}| \right) ,
	%	\end{aligned}
%\end{equation}	
%where $th=\frac{G'\left[ |E_{s1}| \right]|E_{s2}|}{2}$ is judgment threshold and $Q(\cdot)$ is Q-function.

$\ \ $ Under a fixed high-strength electric field $|E_{s1}|$, the average APD voltage of bits zero and one are $G\left[ |E_{s1}|,0 \right]$ and $G\left[ |E_{s1}|,|E_{s2}| \right]$, respectively. The detailed statistical distribution of bits is shown in Fig. \ref{PYDisCase1I14S15}. Thus, the BER can be expressed as
\begin{equation}\label{C1BERCalculationFixSiF}
	\begin{aligned}
		P_{C1}&=\frac{1}{2}\int_{th}^{+\infty}{\frac{1}{\sqrt{2\pi}\sigma _{}}e^{-\frac{\left( x-G\left[ |E_{s1}|,0 \right] \right) ^2}{2\sigma _{}^{2}}}dx}
		\\
		&\ \ \ +\frac{1}{2}\int_{-\infty}^{th}{\frac{1}{\sqrt{2\pi}\sigma _{}}e^{-\frac{\left( x-G\left[ |E_{s1}|,|E_{s2}| \right] \right) ^2}{2\sigma _{}^{2}}}dx}
		\\
		&=Q\left(\left|  \frac{G\left[ |E_{s1}|,|E_{s2}| \right] -G\left[ |E_{s1}|,0 \right] }{2\sigma _{}}\right|  \right) ,
	\end{aligned}
\end{equation}	
where $th=\left| \frac{G\left[ |E_{s1}|,|E_{s2}| \right] -G\left[ |E_{s1}|,0| \right]}{2} \right|$ is detection threshold and $Q\left( x \right) =\frac{1}{\sqrt{2\pi}}\int_x^{+\infty}{e^{-\frac{t^2}{2}}dt}$.

According to the features in (\ref{Boundary}) and the simulation results in Fig. \ref{modelshow2}, as signal intensity $|E_{s1}|$ increases, the response coefficient curve $G[|E_{s1}|,\cdot]$ is flattened. Under high-intensity electric field $\left| E_{s1}\right| $, it can be approximated that $G\left[ |E_{s1}|,|E_{s2}| \right] \approx G\left[ |E_{s1}|,0 \right]$. Thus, we have
%\begin{equation}\label{BoundaryC1BERCalculationFixSaF}
%	\begin{aligned}
	%\lim_{|E_{s1}|\rightarrow +\infty}P_{C1,CELM}&=Q\left( 0 \right) ,
	%	\end{aligned}
%\end{equation}	
\begin{equation}\label{BoundaryC1C1BERCalculationFixSiF}
	\begin{aligned}
		\lim_{|E_{s1}|\rightarrow +\infty}P_{C1}&=Q\left( 0 \right) ,
	\end{aligned}
\end{equation}	
where $Q\left( 0 \right)=0.5$. It also indicates that high-intensity electric field can seriously affect communication performance of signal $E_{s2}$.
%\begin{equation}\label{Boundary}
%	\begin{aligned}
	%%\lim_{|E_1|\rightarrow +\infty}\frac{\partial G\left[ |E_{s1}|,\cdot \right]}{\partial |E_{s1}|}&=0,
	%%\\
	%%\lim_{|E_1|\rightarrow +\infty}\frac{\partial G\left[ \cdot ,\left| E_{s1} \right| \right]}{\partial |E_{s1}|}&=0, and its slope $\frac{\partial G\left[ |E_{s1}|,\cdot \right]}{\partial |E_{s1}|}$
	%%\\
	%\lim_{|E_{1}|\rightarrow +\infty}G\left[ |E_{1}|,\cdot \right] &=0,
	%\\
	%\lim_{|E_{1}|\rightarrow +\infty}G\left[\cdot,|E_{1}| \right] &=0.
	%	\end{aligned}
%\end{equation}	
%Due to normalization joint response coefficient, we have $G[0,0]=1$. 

\subsection{Case 2: The BER of Signal $E_{s1}$ with Constant Intensity $E_{s2}$.}\label{T32}

$\ \ $ Under a fixed low-strength electric field $|E_{s2}|$, the APD mean voltages for bits zero and one are $G\left[ 0,|E_{s2}| \right]$ and $G\left[ |E_{s1}|,|E_{s2}| \right]$, respectively. The detailed statistical distribution of bits is shown in Fig. \ref{PYDisCase2I14S15}. The BER can be expressed as
\begin{equation}\label{C2BERCalculationFixSiF}
	\begin{aligned}
		P_{C2}&=\frac{1}{2}\int_{th}^{+\infty}{\frac{1}{\sqrt{2\pi}\sigma _{}}e^{-\frac{\left( x-G\left[ 0 ,|E_{s2}|\right] \right) ^2}{2\sigma _{}^{2}}}dx}
		\\
		&\ \ \ +\frac{1}{2}\int_{-\infty}^{th}{\frac{1}{\sqrt{2\pi}\sigma _{DC}}e^{-\frac{\left( x-G\left[ |E_{s1}|,|E_{s2}| \right] \right) ^2}{2\sigma _{}^{2}}}dx}
		\\
		&=Q\left( \left| \frac{G\left[ |E_{s1}|,|E_{s2}| \right] -G\left[ 0 ,|E_{s2}|\right] }{2\sigma _{}}\right|  \right) ,
	\end{aligned}
\end{equation}	
where $th=\left| \frac{G\left[ |E_{s1}|,|E_{s2}| \right] -G\left[ 0,|E_{s2}| \right]}{2} \right|$ is detection threshold.

According to the features in (\ref{Boundary}), for signal with higher intensity $|E_{s1}|$, we have 
%\begin{equation}\label{BoundaryC2BERCalculationFixSaF}
%	\begin{aligned}
	%		\lim_{|E_{s1}|\rightarrow +\infty}P_{C2,CELM}&=Q\left( \frac{G\left[ |E_{s2}| \right]}{2\sigma _{DC}} \right) ,
	%	\end{aligned}
%\end{equation}	
\begin{equation}\label{BoundaryC2C1BERCalculationFixSiF}
	\begin{aligned}
		\lim_{|E_{s1}|\rightarrow +\infty}P_{C2}&=Q\left( \frac{G\left[ 0,|E_{s2}| \right] }{2\sigma _{}} \right).
	\end{aligned}
\end{equation}

Under the interference of the fixed low-strength electric field $|E_{s2}|$, increasing the strength of the signal $|E_{s1}|$ will reduce the BER, but the limiting performance depends on noise $\sigma$ and $|E_{s2}|$.

\subsection{Case 3: The BER of Signal $E_{s2}$ under Random OOK Signal $E_{s1}$.}\label{T33}

$\ \ $Under random high-power OOK signal $E_{s1}$, the four APD average voltage combinations 00, 01, 10 and 11 of two users' OOK bits are $G\left[ 0,0 \right]$, $G\left[ 0,|E_{s2}| \right]$, $G\left[ |E_{s1}|,0 \right]$ and $G\left[ |E_{s1}|,|E_{s2}| \right]$, respectively. Without losing generality, let $G\left[ 0,0 \right]>G\left[ 0,|E_{s2}| \right]>G\left[ |E_{s1}|,0 \right]>G\left[ |E_{s1}|,|E_{s2}| \right]$. Then, bits zero and one of signal $E_{s2}$ follow bimodal Gaussian distribution of
$\frac{1}{2}\mathcal{N}\left( G\left[ 0,0 \right] ,\sigma \right) +\frac{1}{2}\mathcal{N}\left( G\left[ |E_{s1}|,0 \right] ,\sigma \right) $ and $\frac{1}{2}\mathcal{N}\left( G\left[ 0,|E_{s2}| \right] ,\sigma \right) +\frac{1}{2}\mathcal{N}\left( G\left[ |E_{s1}|,|E_{s2}| \right] ,\sigma \right) $, respectively. The detailed statistical distribution of bits is shown in Fig. \ref{PYDisCase3I14S15}. Thus, the calculation of BER is shown in (\ref{C3BERCalculationRandomSiF})
%\begin{figure*}
%	\begin{equation}\label{C3BERCalculationRandomSiF}
%		\begin{aligned}
%			P_{C3}&=\frac{1}{4}P_{00}+\frac{1}{4}P_{01}+\frac{1}{4}P_{10}+\frac{1}{4}P_{11}
%			\\
%			&=\frac{1}{4}\int_{th_{1}}^{+\infty}{\frac{1}{\sqrt{2\pi}\sigma _{}}e^{-\frac{\left( x-G\left[ 0,0 \right] \right) ^2}{2\sigma _{}^{2}}}dx}
%			\\
%			&\ \ \ +\frac{1}{4}\int_{-\infty}^{th_{1}}{\frac{1}{\sqrt{2\pi}\sigma _{}}e^{-\frac{\left( x-G\left[ 0,|E_{s2}| \right] \right) ^2}{2\sigma _{}^{2}}}dx}
%			\\
%			&\ \ \ +\frac{1}{4}\int_{th_{2}}^{+\infty}{\frac{1}{\sqrt{2\pi}\sigma _{}}e^{-\frac{\left( x-G\left[ 0,|E_{s2}| \right] \right) ^2}{2\sigma _{}^{2}}}dx}
%			\\
%			&\ \ \ +\frac{1}{4}\int_{-\infty}^{th_{2}}{\frac{1}{\sqrt{2\pi}\sigma _{}}e^{-\frac{\left( x-G\left[ |E_{s1}|,0 \right] \right) ^2}{2\sigma _{}^{2}}}dx}
%			\\
%			&\ \ \ +\frac{1}{4}\int_{th_{3}}^{+\infty}{\frac{1}{\sqrt{2\pi}\sigma _{}}e^{-\frac{\left( x-G\left[ |E_{s1}|,0 \right] \right) ^2}{2\sigma _{}^{2}}}dx}
%			\\
%			&\ \ \ +\frac{1}{4}\int_{-\infty}^{th_{3}}{\frac{1}{\sqrt{2\pi}\sigma _{}}e^{-\frac{\left( x-G\left[ |E_{s1}|,|E_{s2}| \right] \right) ^2}{2\sigma _{}^{2}}}dx}
%			\\
%			&=\frac{1}{2}Q\left( \left| \frac{1 -G\left[ 0,|E_{s2}| \right] }{2\sigma _{}}\right|  \right)
%			\\ 
%			&\ \ \ +\frac{1}{2}Q\left( \left| \frac{G\left[ 0,|E_{s2}| \right] -G\left[ |E_{s1}|,0 \right] }{2\sigma _{}}\right|  \right) 
%			\\
%			&\ \ \ +\frac{1}{2}Q\left(\left|  \frac{G\left[ |E_{s1}|,0 \right] -G\left[ |E_{s1}|,|E_{s2}| \right] |}{2\sigma _{}}\right|  \right), 
%		\end{aligned}
%	\end{equation}	
%\end{figure*}
\begin{figure*}
\begin{equation}\label{C3BERCalculationRandomSiF}
	\begin{aligned}
		P_{C3}&=\frac{1}{4}P_{00}+\frac{1}{4}P_{01}+\frac{1}{4}P_{10}+\frac{1}{4}P_{11}
		\\
		&=\frac{1}{4}\int_{th_{1}}^{+\infty}{\frac{1}{\sqrt{2\pi}\sigma _{}}e^{-\frac{\left( x-G\left[ 0,0 \right] \right) ^2}{2\sigma _{}^{2}}}dx}+\frac{1}{4}\int_{-\infty}^{th_{1}}{\frac{1}{\sqrt{2\pi}\sigma _{}}e^{-\frac{\left( x-G\left[ 0,|E_{s2}| \right] \right) ^2}{2\sigma _{}^{2}}}dx}+\frac{1}{4}\int_{th_{2}}^{+\infty}{\frac{1}{\sqrt{2\pi}\sigma _{}}e^{-\frac{\left( x-G\left[ 0,|E_{s2}| \right] \right) ^2}{2\sigma _{}^{2}}}dx}
		\\
		&\ \ \ +\frac{1}{4}\int_{-\infty}^{th_{2}}{\frac{1}{\sqrt{2\pi}\sigma _{}}e^{-\frac{\left( x-G\left[ |E_{s1}|,0 \right] \right) ^2}{2\sigma _{}^{2}}}dx}+\frac{1}{4}\int_{th_{3}}^{+\infty}{\frac{1}{\sqrt{2\pi}\sigma _{}}e^{-\frac{\left( x-G\left[ |E_{s1}|,0 \right] \right) ^2}{2\sigma _{}^{2}}}dx}+\frac{1}{4}\int_{-\infty}^{th_{3}}{\frac{1}{\sqrt{2\pi}\sigma _{}}e^{-\frac{\left( x-G\left[ |E_{s1}|,|E_{s2}| \right] \right) ^2}{2\sigma _{}^{2}}}dx}
		\\
		&=\frac{1}{2}Q\left( \left| \frac{1 -G\left[ 0,|E_{s2}| \right] }{2\sigma _{}}\right|  \right)+\frac{1}{2}Q\left( \left| \frac{G\left[ 0,|E_{s2}| \right] -G\left[ |E_{s1}|,0 \right] }{2\sigma _{}}\right|  \right)+\frac{1}{2}Q\left(\left|  \frac{G\left[ |E_{s1}|,0 \right] -G\left[ |E_{s1}|,|E_{s2}| \right] |}{2\sigma _{}}\right|  \right), 
	\end{aligned}
\end{equation}	
\end{figure*}
,where we have $G[0,0]=1$. Let $P_{00}$, $P_{01}$, $P_{10}$ and $P_{11}$ be the error probability of each combinations. Let $th_{1}=\left| \frac{1-G\left[ 0,|E_{s2}| \right]}{2} \right|$, $th_{2}=\left| \frac{G\left[ 0,|E_{s2}| \right] -G\left[ |E_{s1}|,0 \right]}{2} \right|$ and $th_{3}=\left| \frac{G\left[ |E_{s1}|,0 \right] -G\left[ |E_{s1}|,|E_{s2}| \right]}{2} \right|$ be the detection thresholds of each combination.

According to the features in (\ref{Boundary}), for random signal with higher intensity $|E_{s1}|$, we have
\begin{equation}\label{BoundaryC2BERCalculationFixSaFSiF}
	\begin{aligned}
		\lim_{|E_{s1}|\rightarrow +\infty}P_{C3}&=\frac{1}{4}+\frac{1}{2}Q\left(\left|  \frac{1 -G\left[ 0,|E_{s2}| \right] }{2\sigma _{}}\right|  \right)
		\\ 
		&\ \ \ +\frac{1}{2}Q\left( \left| \frac{G\left[ 0,|E_{s2}| \right] -0 }{2\sigma _{}}\right|  \right) .
	\end{aligned}
\end{equation}	

The high intensity signal $|E_{s1}|$ would affect the reception of $E_{s2}$ with $P_{C2}>0.25$ based on (\ref{BoundaryC2BERCalculationFixSaFSiF}).
In addition, though the two signal couple the different atomic energy levels, due to the presence of specific strength $|E_{s1}|$ satisfying $G\left[ 0,|E_{s2}| \right] \approx G\left[ |E_{s1}|,0 \right] $, the communication performance of signal $E_{s2}$ would still deteriorate, which is given by
\begin{equation}\label{C2BERCalculationRandomSiFPeak}
	\begin{aligned}	
		P_{C3}&=\frac{1}{4}+\frac{1}{2}Q\left( \left| \frac{1 -G\left[ 0,|E_{s2}| \right] }{2\sigma _{}}\right|  \right)
		\\
		&\ \ \ +\frac{1}{2}Q\left( \left| \frac{G\left[ |E_{s1}|,0 \right] -G\left[ |E_{s1}|,|E_{s2}| \right] }{2\sigma _{}}\right| \right)
		\\
		&>0.25.
	\end{aligned}
\end{equation}	
\subsection{Case 4: The BER of Signal $E_{s1}$ under Random OOK Signal $E_{s2}$.}\label{T34}
%
%
%Under CELM and two random OOK signal, the symbols of two signals 00, 01, 10 and 11 follow the complex Gaussian distribution with mean $G[0]$, $G[|E_{s2}|]$, $G[|E_{s1}|]$ and $G[|E_{s1}|]+iG'[|E_{s1}|]|E_{s2}|$, respectively. Due to symbols 00 and 01 have the same symbol of signal $E_{s1}$, the BER between 00 and 01 can be seen as 0. And symbols 10 and 11 are similar. Thus, the BER of simplified CELM only consider symbols 01 and 10, which is given by $P_{C3,01}$ and $P_{C3,10}$, respectively. Under Gaussian noise, the BER of signal $E_{s1}$ can be approximately expressed as
%
%\begin{equation}\label{C4BERCalculationRandomSaF}
%	\begin{aligned}	
	%P_{C4,CELM}&=\frac{1}{4}P_{CELM,00}+\frac{1}{4}P_{CELM,01}
	%\\
	%&\ \ \ +\frac{1}{4}P_{CELM,10}+\frac{1}{4}P_{CELM,11}
	%\\
	%&=\frac{1}{4}\int_{th_{DC1}}^{+\infty}{\frac{1}{\sqrt{2\pi}\sigma _{DC}}e^{-\frac{\left( x-G\left[ |E_{s1}| \right] \right) ^2}{2\sigma _{DC}^{2}}}dx}
	%\\
	%&\ \ \ +\frac{1}{4}\int_{-\infty}^{th_{DC1}}{\frac{1}{\sqrt{2\pi}\sigma _{DC}}e^{-\frac{\left( x-G\left[ |E_{s2}| \right] \right) ^2}{2\sigma _{AC}^{2}}}dx}
	%\\
	%&=\frac{1}{2}Q\left( \frac{|G\left[ |E_{s2}| \right] -G\left[ |E_{s1}| \right] |}{2\sigma _{DC}} \right) ,
	%	\end{aligned}
%\end{equation}
%where $P_{CELM,00}=0$, $P_{CELM,11}=0$ and $th_{DC1}=\left| \frac{G\left[ |E_{s2}| \right] -G\left[ |E_{s1}| \right]}{2} \right|$.

$\ \ $Under random OOK signal $E_{s2}$, consider without losing generality, let $G\left[ 0,0 \right]>G\left[ 0,|E_{s2}| \right]>G\left[ |E_{s1}|,0 \right]>G\left[ |E_{s1}|,|E_{s2}| \right]$, bits zero and one of $E_{s1}$ follow bimodal Gaussian distribution, which are given by
\begin{equation}
	\begin{aligned}	
		&\text{Bit Zero}\sim \frac{1}{2}\mathcal{N}\left( G\left[ 0,0 \right] ,\sigma \right) +\frac{1}{2}\mathcal{N}\left( G\left[ 0,|E_{s2}| \right] ,\sigma \right)
		\\
		&\text{Bit One}\sim \frac{1}{2}\mathcal{N}\left( G\left[ |E_{s1}|,0 \right] ,\sigma \right)+\frac{1}{2}\mathcal{N}\left( G\left[ |E_{s1}|,|E_{s2}| \right] ,\sigma \right).
	\end{aligned}
\end{equation}

The simplified model only consider the BER between combinations 01 and 10, which are given by $P_{01}$ and $P_{10}$, respectively. The detailed statistical distribution of bits is shown in Fig. \ref{PYDisCase4I14S15}. The BER of signal $E_{s1}$ can be approximated as

\begin{equation}\label{C4BERCalculationRandomSiF}
	\begin{aligned}	
		P_{C4}&=\frac{1}{4}P_{00}+\frac{1}{4}P_{01}+\frac{1}{4}P_{10}+\frac{1}{4}P_{11}
		\\
		&=\frac{1}{4}\int_{th_{2}}^{+\infty}{\frac{1}{\sqrt{2\pi}\sigma _{}}e^{-\frac{\left( x-G\left[ 0,|E_{s2}| \right] \right) ^2}{2\sigma _{}^{2}}}dx}
		\\
		&\ \ \ +\frac{1}{4}\int_{-\infty}^{th_{2}}{\frac{1}{\sqrt{2\pi}\sigma _{}}e^{-\frac{\left( x-G\left[ |E_{s1}|,0 \right] \right) ^2}{2\sigma _{}^{2}}}dx}
		\\
		&=\frac{1}{2}Q\left( \left| \frac{G\left[ 0,|E_{s2}| \right] -G\left[ |E_{s1}|,0 \right] }{2\sigma _{}} \right| \right) ,
	\end{aligned}
\end{equation}
where $P_{00}=0$, $P_{11}=0$ and $th_{2}=\left| \frac{G\left[ 0,|E_{s2}| \right] -G\left[ |E_{s1}|,0 \right]}{2} \right|$.

According to the features in (\ref{Boundary}), for random signal with higher electric field intensity $|E_{s1}|$, we have
\begin{equation}\label{BoundaryC3BERCalculationFixSaFSiF}
	\begin{aligned}
		\lim_{|E_{s1}|\rightarrow +\infty}P_{C4}&=\frac{1}{2}Q\left( \left| \frac{G\left[ 0,|E_{s2}| \right]}{2\sigma _{}}\right|  \right) .
	\end{aligned}
\end{equation}	

Similar to Case 2, under the interference of random OOK signal $E_{s2}$, increasing power of $E_{s1}$ will reduce the BER of signal $E_{s1}$, but the limiting performance depends on noise $\sigma$ and $|E_{s2}|$.

\subsection{SER Performance}

$\ \ $ Under two random OOK signal from two users, the APD average voltage of the four combinations 00, 01, 10 are $G\left[ 0,0 \right]$, $G\left[ 0,|E_{s2}| \right]$, $G\left[ |E_{s1}|,0 \right]$ and $G\left[ |E_{s1}|,|E_{s2}| \right]$, respectively. Without losing generality, let $G\left[ 0,0 \right]>G\left[ 0,|E_{s2}| \right]>G\left[ |E_{s1}|,0 \right]>G\left[ |E_{s1}|,|E_{s2}| \right]$. Based on the statistical distribution of each voltage combinations shown in Fig.\ref{Dis}. Moreover, the SER can be expressed as

\begin{equation}\label{}
	\begin{aligned}
		P_{SER}&=\frac{1}{4}P_{00}+\frac{1}{4}P_{01}+\frac{1}{4}P_{10}+\frac{1}{4}P_{11}
		\\
		&=\frac{1}{2}Q\left( \left| \frac{1 -G\left[ 0,|E_{s2}| \right] }{2\sigma _{}}\right|  \right)
		\\ 
		&\ \ \ +\frac{1}{2}Q\left( \left| \frac{G\left[ 0,|E_{s2}| \right] -G\left[ |E_{s1}|,0 \right] }{2\sigma _{}}\right|  \right) 
		\\
		&\ \ \ +\frac{1}{2}Q\left(\left|  \frac{G\left[ |E_{s1}|,0 \right] -G\left[ |E_{s1}|,|E_{s2}| \right] |}{2\sigma _{}}\right|  \right), 
	\end{aligned}
\end{equation}	
where $G[0,0]=1$; $P_{00}$, $P_{01}$, $P_{10}$ and $P_{11}$ are the error probability of each combinations.
Approximatively, we can consider $P_{SER}$ as equal to $P_{C3}$ due to the fact each symbol contain one bit for signal $E_{s2}$.

\section{Experimental Results}\label{T4}

$\ \ $ In the experiment, we utilize a 5 cm-long Rubidium vapor cell at room temperature, filled with ground-state atoms at an approximate density of $N_0=1.29\times10 ^{16} \ \text{m}^{-3}$. The probe ($\lambda_p=780 \ \text{nm}$) and coupling laser ($\lambda_c=480 \  \text{nm}$) beams excite the atoms from state $5S_{1/2}$ to state $5P_{3/2} $ and state $5P_{3/2} $ to state $53D_{5/2}$, respectively. For RF signals, we consider Rydberg state $53D_{5/2}$. The RF frequency $f_1=14.233\ \text{GHz}$ and $f_2=15.58\ \text{GHz}$ resonantly drive the Rydberg transition from state $53D_{5/2}$ to $54P_{3/2}$ and from state $53D_{5/2}$ to $52F_{7/2}$, respectively. Let $e$ and $a_0$ denote the electronic charge and Bohr radius, respectively. The projection of total orbital angular momentum of $53D_{5/2}$, $54P_{3/2}$ and $52F_{7/2}$ are $1/2$. The dipole moment from state $53D_{5/2}$ to $54P_{3/2}$ and from state $53D_{5/2}$ to $52F_{7/2}$ are $\mu_1=1774.82ea_0$ and $\mu_2=1778.72ea_0$, respectively. At the transmitter, the two OOK signal in RF source A and B are sent by one horn antenna. For the probe beam, the $1/e^2$ beam radius is 0.35 mm. For the coupling beam, the $1/e^2$ beam radius is 0.24 mm. We set probe laser and coupling laser power to 4.14 $\mu$W and 119 mW, respectively. The Rabi frequencies of probe laser and coupling laser are $\Omega_{p}/2\pi=1.99$ MHz and $\Omega_{c}/2\pi=4.51$ MHz, respectively. The output voltage of APD with 400 M bandwidth is sampled directly by a high-speed data collector at the maximum rate of 100 M/s. In order to reduce noise, a $48$ MHz low pass filter was adopted before the APD voltage data was collected, and the sampled data are stored for further offline processing. Both lasers have no detuning. All the data related to Rb atoms is calculated by Alkali Rydberg Calculator (ARC) package\cite{vsibalic2017arc}.

\subsection{Joint Response Coefficient $G$}\label{T41}

\begin{figure}[htbp]
	\centering  %图片全局居中
	\includegraphics[width=0.4\textwidth]{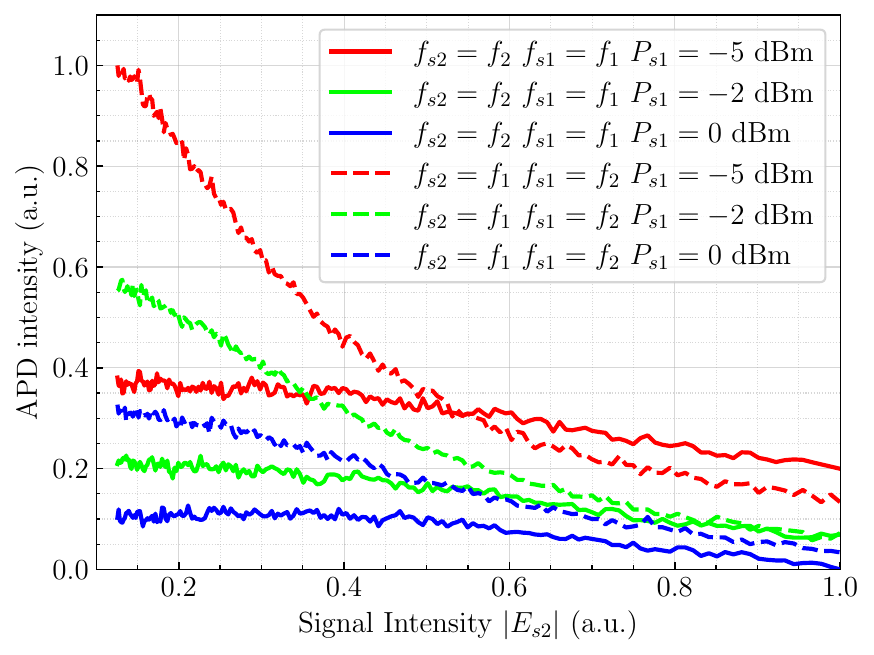}
	\caption{Normalized APD voltage with respect to signal intensity. The solid line: signal $E_{s1}$ with frequency $f_1$ couples dipole moment $\mu_{1}$ and signal $E_{s2}$ with frequency $f_2$ couples dipole moment $\mu_{2}$. The dash line: signal $|E_{s1}|$ with frequency $f_2$ couples dipole moment $\mu_{2}$ and signal $|E_{s2}|$ with frequency $f_1$ couples dipole moment $\mu_{1}$.}
	\label{I15}
\end{figure}

$\ \ $ In the experiment, we test the normalized APD intensity under different signal power and the results shown in Fig. \ref{I15}. The instability of the output power of the RF source and non ideal antennas and feeders cause small scale fluctuations in the data. In large scale, the experimental results and theoretical analysis displayed in Section \ref{T2} fit well in terms of concavity and convexity. The red, green and blue lines correspond to three RF intensities $P_{s1}=-5$ dBm, $P_{s1}=-10$ dBm and $P_{s1}=0$ dBm, respectively. High electric field strength $|E_{s1}|$ flatten the APD voltage curve to reduce the APD output. The experimental results of Fig. \ref{I15} match well with the theoretical analysis in Fig. \ref{modelshow2}. 

In addition, experimental results indicate that Rydberg sensors have better reception capability for $f_1=14.233\ \text{GHz}$ signals compared to $f_1=15.58\ \text{GHz}$ signals. Under the same color, the solid lines are flatter than the dashed lines in Fig. \ref{I15}. And in later experimental results shown in Sec. \ref{T42} and Sec. \ref{T43}, the signal $E_{s1}$ or $E_{s2}$ coupling state $52F_{7/2}$ require higher power to achieve the same effect. 

%\begin{figure}[htbp]
%	\centering  %图片全局居中
%	\includegraphics[width=0.4\textwidth]{fig2//PYmuRF.pdf}
%	\caption{Partial available atomic energy levels of Rb atoms $53D_{5/2}$.}
%	\label{muRF}
%\end{figure}

\subsection{Waveform and Statistical Distribution}\label{T42}
%In Fig. \ref{WaveForm}, we fix the E-filed intensity of single $|E_{s2}|= 0.4$ mV/cm and show the effect of two models on time-domain waveform. In Fig. \ref{WaveFormI14S14}, we consider two signals couple the same atomic energy level from state $53D_{5/2}$ to $54P_{3/2}$. The frequency $f_{s1}$ of signal $E_{s1}$ is equal to the resonance frequency, i.e., $f_{s1}=f_1$, and there is a small difference $\Delta f=100$ KHz in frequency between signal $E_{s2}$ and $E_{s1}$, i.e., $f_{s2}=f_1+\Delta f$. In Fig. \ref{WaveFormI15S15}, we adopt another atomic energy level from state $53D_{5/2}$ to $52F_{5/2}$. Then, we consider the two RF signals with the resonance frequencies couple the different levels. In Fig. \ref{WaveFormI14S15}, we set $f_{s1}=f_2$ and $f_{s2}=f_1$. In Fig. \ref{WaveFormI15S14}, $f_{s1}=f_1$ and $f_{s2}=f_2$. In Fig. \ref{WaveFormI14S14} and Fig. \ref{WaveFormI15S15}, the AC part with frequency $\Delta f$ would be blocked with high power $E_{s1}$. And cmpared to CELM, in Fig. \ref{WaveFormI14S15} and Fig. \ref{WaveFormI15S14}, only DC part can be collected to distinguish symbols.

\begin{figure}[htbp]
	\centering  %图片全局居中
	\includegraphics[width=0.4\textwidth]{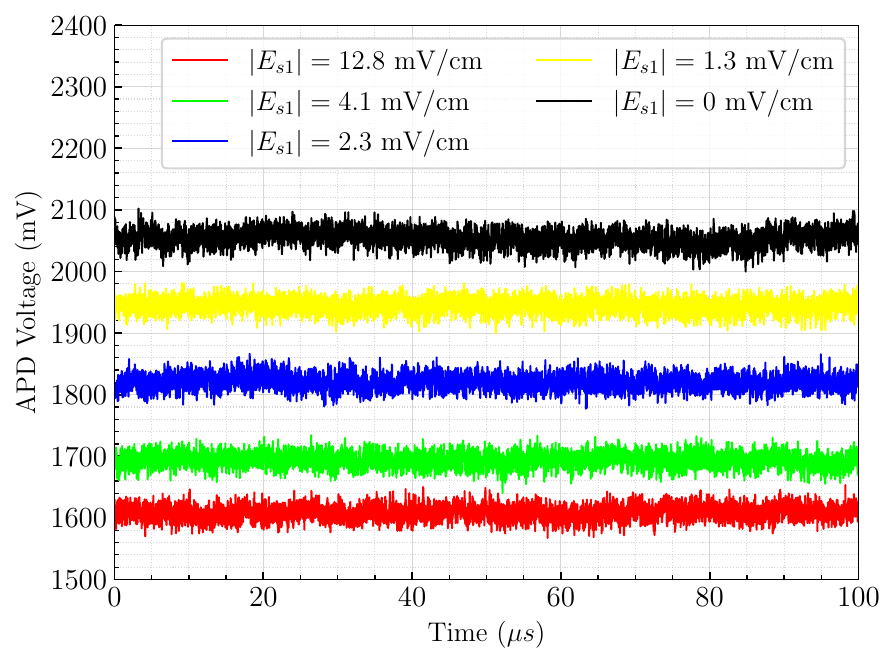}
	\caption{Waveform of $f_{s1}=f_1$ and $f_{s2}=f_{2}$.}
	\label{WaveFormI14S15}
\end{figure}

\begin{figure}[htbp]
	\centering  %图片全局居中
	\includegraphics[width=0.4\textwidth]{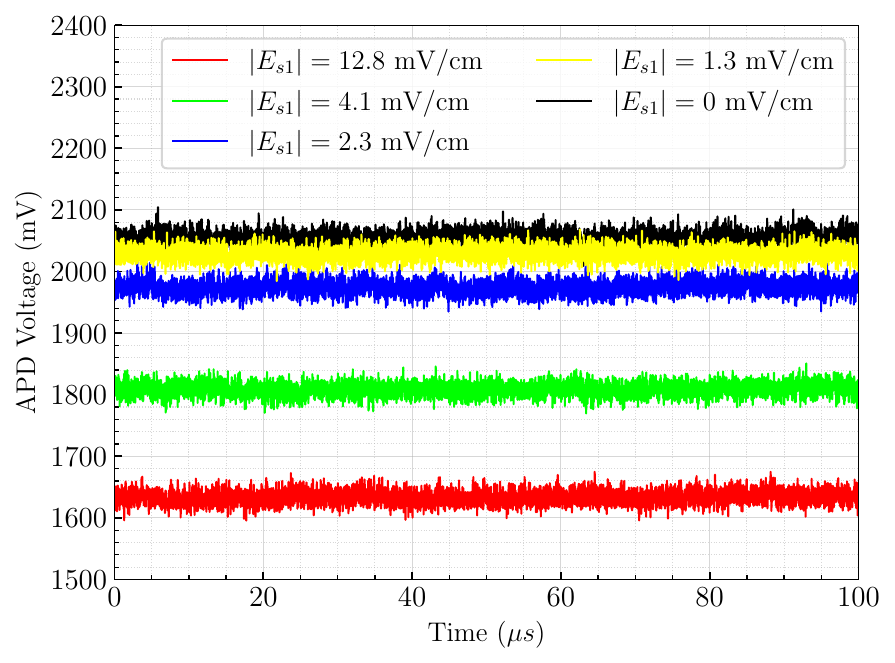}
	\caption{Waveform of $f_{s1}=f_2$ and $f_{s2}=f_{1}$.}
	\label{WaveFormI15S14}
\end{figure}

$\ \ $ In Fig. \ref{WaveFormI14S15} and Fig. \ref{WaveFormI15S14}, we fix electric field intensity $|E_{s2}|= 0.4$ mV/cm and show the effect of sensing ability on time-domain waveform. In Fig. \ref{WaveFormI14S15}, the two signals $E_{s1}$ and $E_{s2}$ couple atomic energy levels from $53D_{5/2}$ to $54P_{3/2}$ and from $53D_{5/2}$ to $52F_{7/2}$, respectively. In Fig. \ref{WaveFormI15S14}, we exchange parameter settings. Due to the stronger sensing ability towards frequency $f_1$, the APD voltage is approximately equal to 1800 mV under the strength of 2.3 mV/cm electric field in Fig. \ref{WaveFormI14S15}. In contrast,  in Fig. \ref{WaveFormI15S14}, in order to achieve the same APD output, a stronger electric field is required for signal coupling transition from $53D_{5/2}$ to $52F_{7/2}$. Regardless of the magnitude of the dipole moment and non ideal cell, in a strong electric field, the APD output will eventually saturate to a fixed value $V_s$, approximately 1600 mV for our experimental setup.

The detailed statistical distribution of each voltage combinations under four cases is shown in Fig. \ref{IQI14S15} and Fig. \ref{Dis}.

\subsection{BER Performance}\label{T43}

%\begin{figure}[htbp]
%	\centering  %图片全局居中
%	\includegraphics[width=0.4\textwidth]{fig2//PYIQplotI14S14.pdf}
%	\caption{The statistical distribution of four states under CELM. We set $E_{s1}=1.8$ mV/cm, $E_{s2}=1.0$ mV/cm, $f_{s1}=f_1$, and $f_{s2}=f_1+ \Delta f$}
%	\label{IQI14S14}
%\end{figure}

%\begin{figure}[htbp]
%	\centering  %图片全局居中
%	\includegraphics[width=0.4\textwidth]{fig2//PYBERSimulationSaFandSiF.pdf}
%	\caption{Calculated BER with respect to signal power $|E_{s1}|$}
%	\label{BERSimulation}
%\end{figure}
%\begin{figure}[htbp]
%	\centering  %图片全局居中
%	\includegraphics[width=0.4\textwidth]{fig2//PYBER.pdf}
%	\caption{Experimental BER with respect to signal power $|E_{s1}|$}
%	\label{BER}
%\end{figure}

\begin{figure*}
	\centering
	\subfigure[]{
		\begin{minipage}[t]{0.5\linewidth}
			\centering
			\includegraphics[width=3 in]{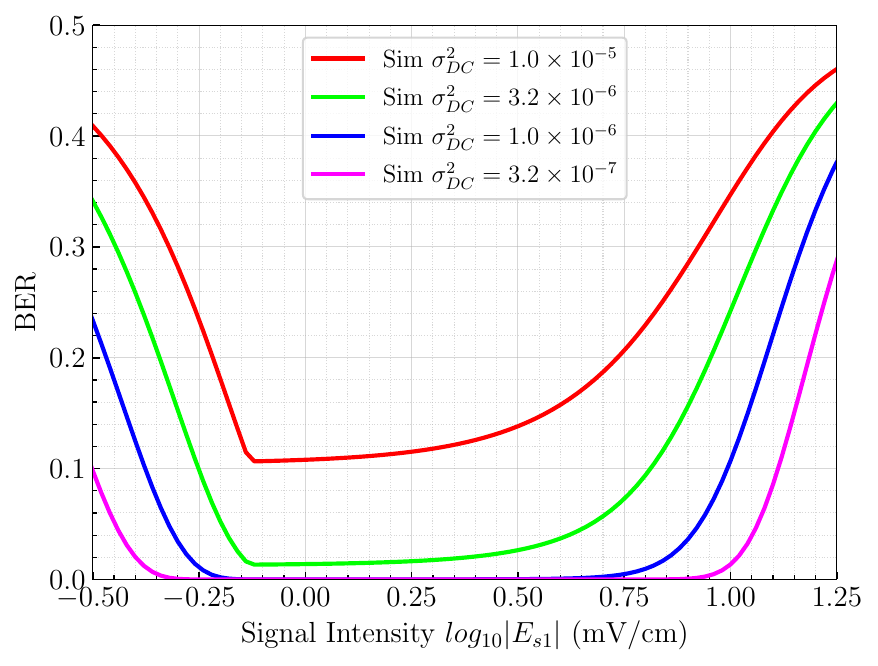}
			%\caption{Symbol switching from zero to zero}
			\label{PYBERSimulation_Case1_SiF}
		\end{minipage}%
	}%
	\subfigure[]{
		\begin{minipage}[t]{0.5\linewidth}
			\centering
			\includegraphics[width=3 in]{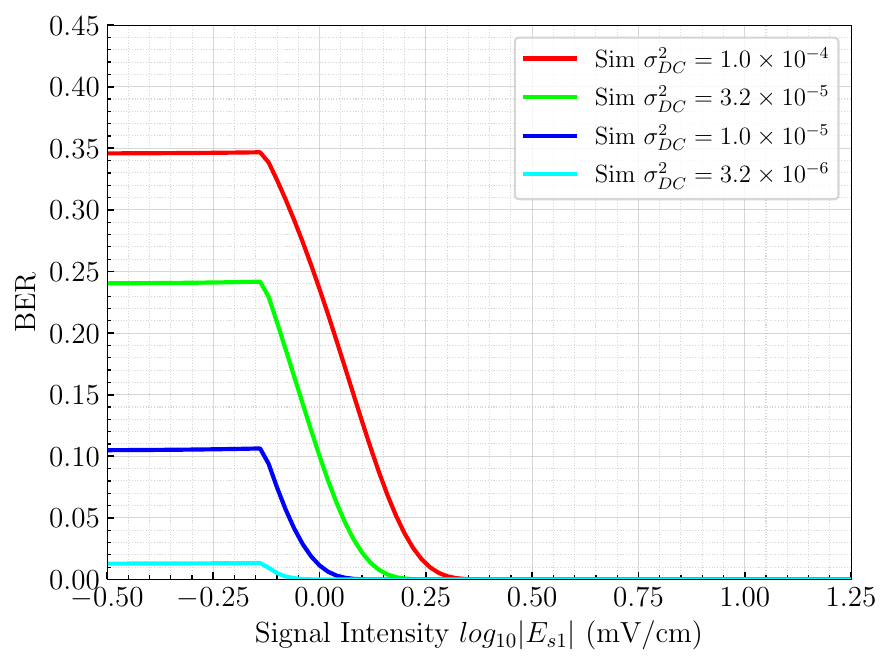}
			\label{PYBERSimulation_Case2_SiF}
			%\caption{Symbol switching from zero to one}
		\end{minipage}%
	}%
	
	\subfigure[]{
		\begin{minipage}[t]{0.5\linewidth}
			\centering
			\includegraphics[width=3in]{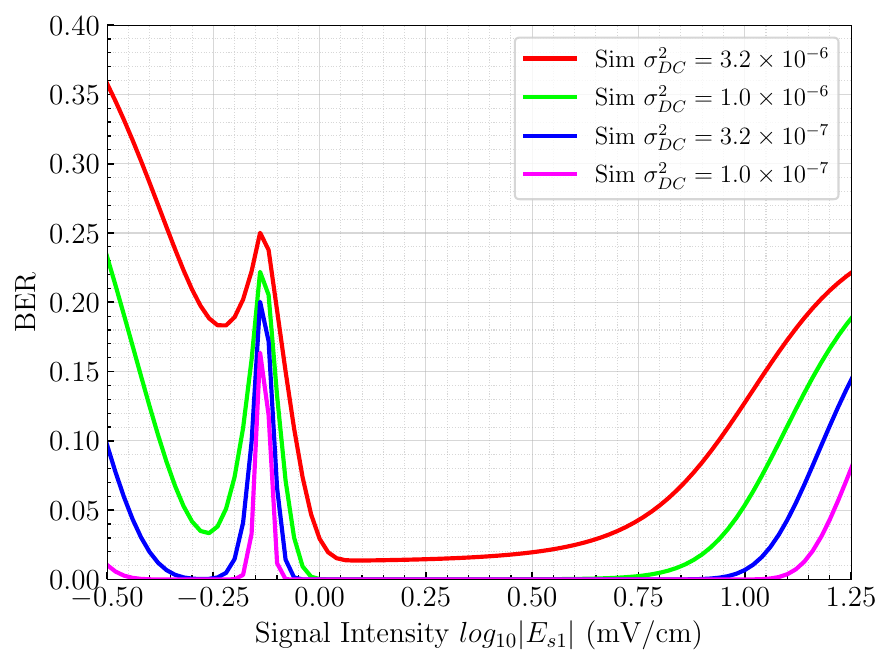}
			\label{PYBERSimulation_Case3_SiF}
			%\caption{Symbol switching from one to zero}
		\end{minipage}
	}%
	\subfigure[]{
		\begin{minipage}[t]{0.5\linewidth}
			\centering
			\includegraphics[width=3in]{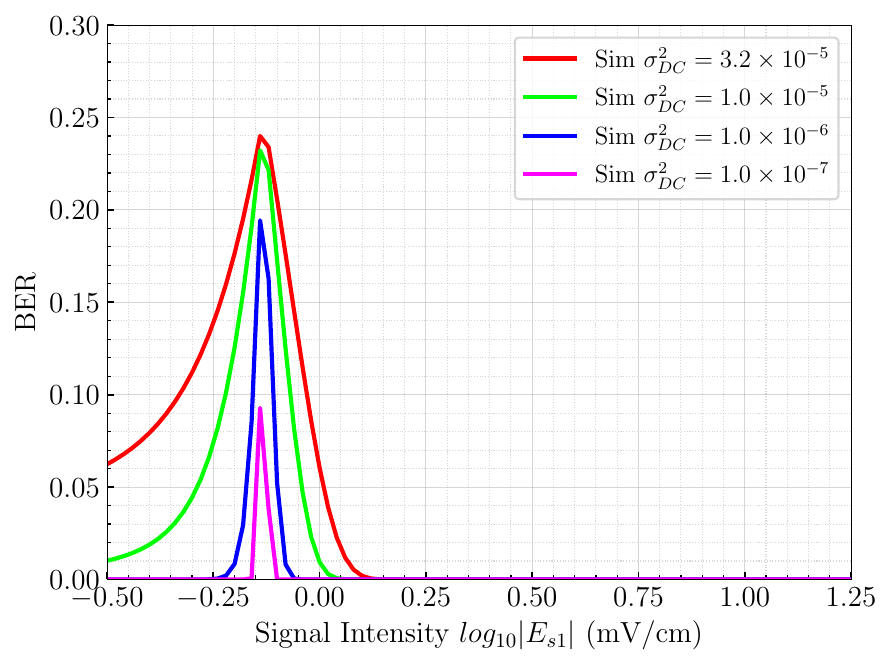}
			\label{PYBERSimulation_Case4_SiF}
			%\caption{Symbol switching from one to one}
		\end{minipage}
	}%
	\centering
	\caption{Simulated BER performance with respect to signal intensity $|E_{s1}|$. (a) Case 1. (b) Case 2. (c) Case 3 (d) Case 4.}
	\label{BERSimulation}
\end{figure*}

\begin{figure*}
	\centering
	\subfigure[]{
		\begin{minipage}[t]{0.5\linewidth}
			\centering
			\includegraphics[width=3 in]{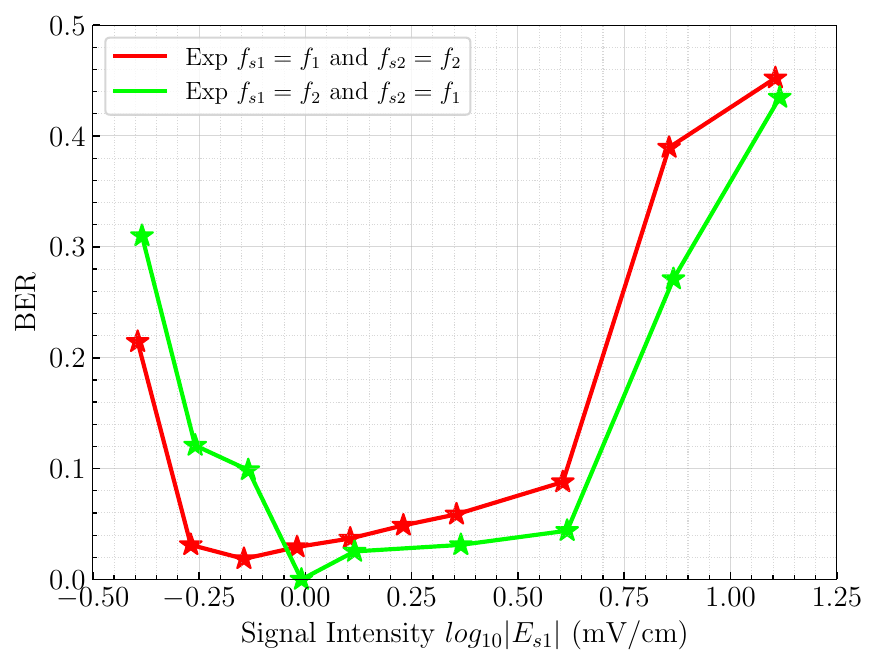}
			%\caption{Symbol switching from zero to zero}
			\label{PYBER_Case1_SiF}
		\end{minipage}%
	}%
	\subfigure[]{
		\begin{minipage}[t]{0.5\linewidth}
			\centering
			\includegraphics[width=3 in]{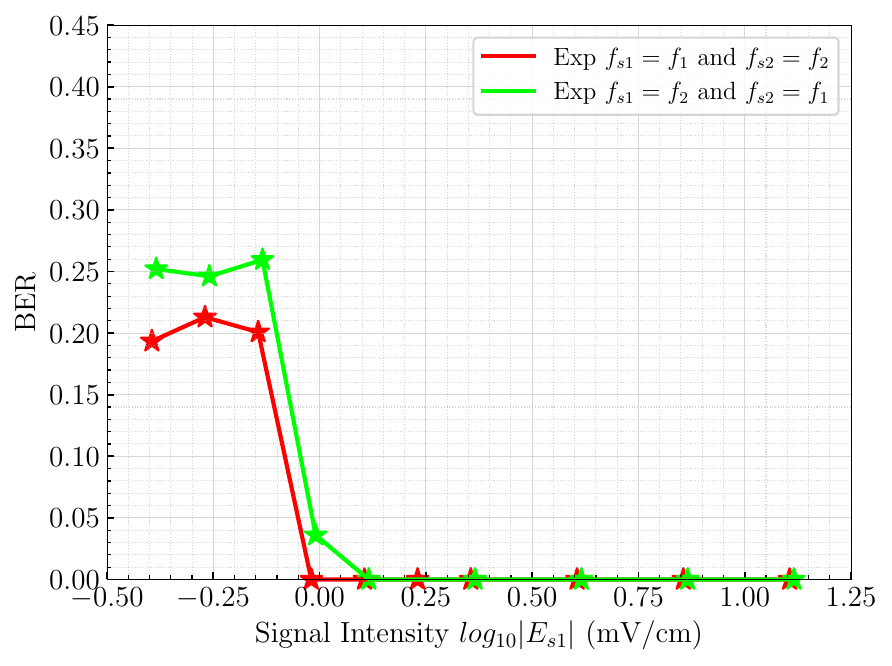}
			\label{PYBER_Case2_SiF}
			%\caption{Symbol switching from zero to one}
		\end{minipage}%
	}%
	
	\subfigure[]{
		\begin{minipage}[t]{0.5\linewidth}
			\centering
			\includegraphics[width=3in]{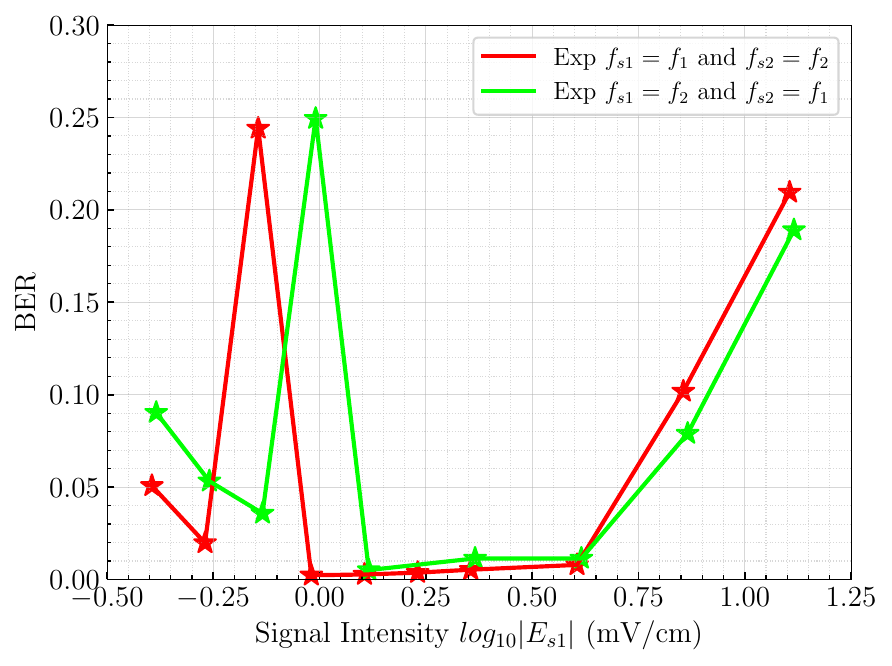}
			\label{PYBER_Case3_SiF}
			%\caption{Symbol switching from one to zero}
		\end{minipage}
	}%
	\subfigure[]{
		\begin{minipage}[t]{0.5\linewidth}
			\centering
			\includegraphics[width=3in]{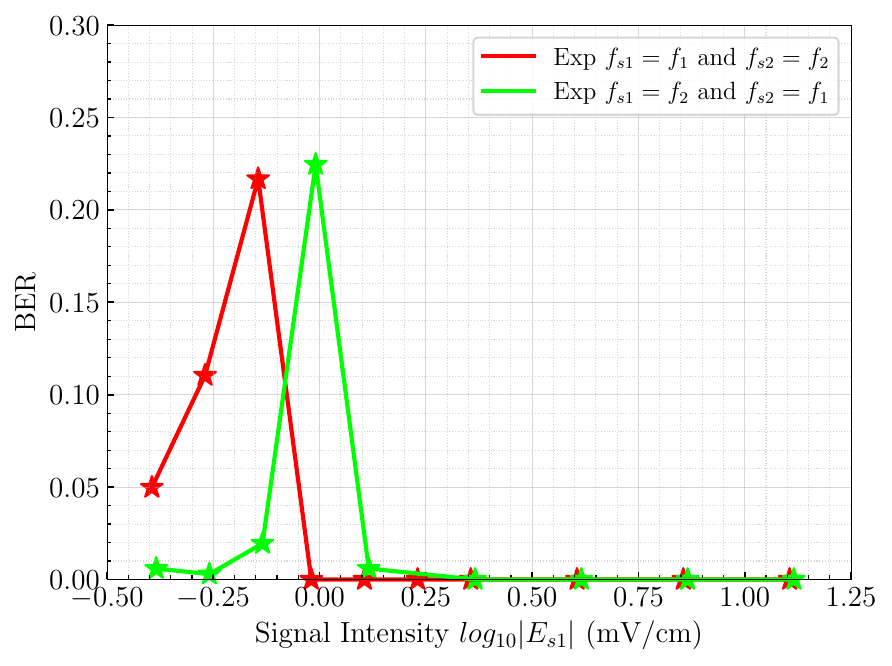}
			\label{PYBER_Case4_SiF}
			%\caption{Symbol switching from one to one}
		\end{minipage}
	}%
	\centering
	\caption{Experimental BER performance with respect to signal intensity $|E_{s1}|$. (a) Case 1. (b) Case 2. (c) Case 3. (d) Case 4.}
	\label{BER}
\end{figure*}

$\ \ $ In Fig. \ref{BERSimulation}, we test the joint response coefficient $G$ in the experiment, and calculate the BER under different noise variances $\sigma_{}$ according to the analysis in Sec. \ref{T4}. In the figures, we fix $|E_{s2}|= 0.4$ mV/cm and show the relationship between BER and electric field intensity $|E_{s1}|$. In Fig. \ref{BER}, the BER results are based on maximum likelihood of the measured random sequence. In the experiment, we randomly generate two synchronized OOK random sequences with symbol period $T_s = 100\ \mu$s, where each sequence contains $2^{11}$ symbols. Because the rising and falling edges caused by symbol switching affect the decision of symbols, we ignore the APD voltage sampled in the first 20 $\mu$s and adopt the samples from 20 $\mu$s to 100 $\mu$s for symbol detection.

For Case 1, the BER decreases with $E_{s1}$ as $\text{log}_{10}|E_{s1}|<0$ mV/cm since the atomic sensors operate in work area of Fig. \ref{modelshow2}, and increases as $\text{log}_{10}|E_{s1}|>0.4$ mV/cm, since it enters the blocked area under higher power $E_{s1}$. For Case 2, the BER decreases for higher power signal $E_{s1}$, and the flat area as $\text{log}_{10}|E_{s1}|<0$ mV/cm can be explained as $G\left[|E_{s1}|,|E_{s2}| \right] \approx G\left[0,|E_{s2}|\right]$ for smaller $|E_{s1}|$. For Case 3, the BER increases as $\text{log}_{10}|E_{s1}|>0.6$ mV/cm due to the interference from higher power $E_{s1}$, i.e., the BER analysis in (\ref{BoundaryC2BERCalculationFixSaFSiF}). It increases and reaches to the peak as $\text{log}_{10}|E_{s1}|<0$ mV/cm due the overlap of statistical distributions between bit combinations 01 and 10, i.e., the analysis in (\ref{C2BERCalculationRandomSiFPeak}). In a word, the peak in BER can be explained as
$G\left[0,|E_{s2}| \right] \approx G\left[ |E_{s1}|,0\right]$ where similar APD outputs cause worse performance. The same reason has also caused the peak in Case 4.

\begin{figure}[htbp]
	\centering  %图片全局居中
	\includegraphics[width=0.4\textwidth]{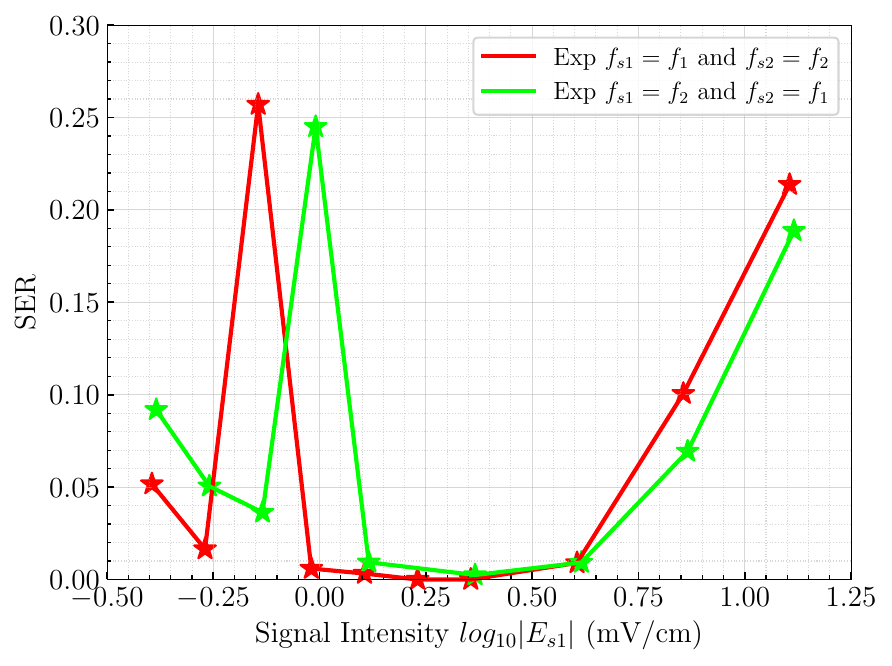}
	\caption{SER performance with respect to signal intensity $|E_{s1}|$.}
	\label{SER_SiF}
\end{figure}

The delay in Fig. \ref{BER} between red and green lines is due to the different sensing ability of two frequencies $f_1$ and $f_2$, which also causes the delay for peaks under Case 3 and Case 4. In Fig. \ref{SER_SiF}, the SER performance shows the same trend as the BER in Case 2 due to the same reasons. The BER of Case 3 is similar to SER due to the fact each symbol contain one bit for signal $E_{s2}$.

\section{Conclusion}\label{T5}
$\ \ $The article investigates the mutual interference for two signals coupling two different energy levels received by a Rydberg atomic sensor. Due to the nonlinearity of receivers, the received signal of atomic sensors corresponding to different energy levels will be downconverted to the baseband simultaneously, resulting in multi-user interference. We have characterized the joint response coefficient and analyzed the detection error rate. We have also conducted experiments to validate the theoretical analysis.

%\newpage

\appendix
\section{Physical Model}\label{A}
\subsection{Atomic Energy Level Model}\label{A1}
$\ \ $ Let $\Delta _p$, $\Delta _c$, $\Delta _{s1}$ and $\Delta _{s2}$ denote the detunings for the probe laser, couple
laser and two RF electric field, respectively. Let $\Omega _p$, $\Omega _c$, $\Omega _{s1}$ and $\Omega _{s2}$ denote Rabi frequencies associated with the probe laser, couple laser and two RF electric field, respectively. We have
\begin{equation}
	\begin{aligned}
		\Omega _{p}&=|E_{p}|\frac{\mu _{p}}{\hbar },\\
		\Omega _{c}&=|E_{c}|\frac{\mu _{c}}{\hbar },\\
		\Omega _{s1}&=|E_{s1}|\frac{\mu _{1}}{\hbar },\\
		\Omega _{s2}&=|E_{s2}|\frac{\mu _{2}}{\hbar },
	\end{aligned}
\end{equation}	
where $|E_{p}|$ and $|E_{c}|$ are the magnitudes of the electric-field of the probe laser, coupling laser, respectively. Let $\mu _{p}$, $\mu _{c}$, $\mu _{1}$ and $\mu _{2}$ denote the atomic dipole moments corresponding to $\left| 1 \right> -\left| 2 \right> $, $\left| 2 \right> -\left| 3 \right> $, $\left| 3 \right> -\left| 4 \right>$ and $\left| 3 \right> -\left| 5 \right>$, respectively. Assume that the decay rates are $\Gamma_{2}$ for state $\left| 2 \right>$, $\Gamma_{3}$ for state $\left| 3 \right>$, $\Gamma_{4}$ for state $\left| 4 \right>$, and $\Gamma_{5}$ for state $\left| 5 \right>$\cite{holloway2017electric}. The corresponding Hamiltonian in the rotating frame is given by
\cite{you2023exclusive}
\begin{equation}
	\begin{aligned}
		H=\frac{{\hbar }}{2}\left[ \begin{matrix}
			0&		\Omega _p&		0&		0&		0\\
			\Omega _p&		0&		\Omega _c&		0&		0\\
			0&		\Omega _c&		0&		\Omega _{s1}&		\Omega_{s2}\\
			0&		0&		\Omega _{s1}&		0&		0\\
			0&		0&		\Omega_{s2}&		0&		0\\
		\end{matrix} \right] .
	\end{aligned}
\end{equation}

The Lindblad operator $\mathcal{L}$ is shown in (\ref{LDoubleModelCase2}).

\begin{figure*}[!t]
	\begin{equation}\label{LDoubleModelCase2}
		\begin{aligned}
			\mathcal{L}=\left[ \begin{matrix}
				\Gamma _2\rho _{22}&		-\gamma _{12}\rho _{12}&		-\gamma _{13}\rho _{13}&		-\gamma _{14}\rho _{14}&		-\gamma _{15}\rho _{15}\\
				-\gamma _{21}\rho _{21}&		-\Gamma _2\rho _{22}+\Gamma _3\rho _{33}&		-\gamma _{23}\rho _{23}&		-\gamma _{24}\rho _{24}&		-\gamma _{25}\rho _{25}\\
				-\gamma _{31}\rho _{31}&		-\gamma _{32}\rho _{32}&		-\Gamma _3\rho _{33}+\Gamma _4\rho _{44}+\Gamma _5\rho _{55}&		-\gamma _{34}\rho _{34}&		-\gamma _{35}\rho _{35}\\
				-\gamma _{41}\rho _{41}&		-\gamma _{42}\rho _{42}&		-\gamma _{43}\rho _{43}&		-\Gamma _4\rho _{44}&		-\gamma _{45}\rho _{45}\\
				-\gamma _{51}\rho _{51}&		-\gamma _{52}\rho _{52}&		-\gamma _{53}\rho _{53}&		-\gamma _{54}\rho _{54}&		-\Gamma _5\rho _{55}\\
			\end{matrix} \right] ,
		\end{aligned}
	\end{equation}
	{\noindent} \rule[-1pt]{18cm}{0.05em}
\end{figure*}

\subsection{Laser Model}\label{A2}

$\ \ $ The relationship between transmittance $T_p$ of the atomic
medium and the linear susceptibility is given by
\begin{equation}
	\begin{aligned}
		T_p=e^{\frac{2N_0\mu _{p}^{2}kL}{\epsilon {\hbar }\Omega _p}\Im \left( \rho _{21}^{SS} \right)},
	\end{aligned}
\end{equation}
where $N_0$ is the total density of atoms, $\epsilon$ is the permittivity in vacuum, $L$ denotes the length of atomic cell, $k=2\pi/\lambda_p$ is the wave vector of the probe laser, and $\Im (\rho_{21}^{SS})$ characterizes the imaginary part of $\rho_{21}$.

In addition, in the thermal atom model, due to the Doppler effect caused by the atomic thermal motion, $\rho_{21D}$ can be expressed as\cite{holloway2017electric}
\begin{equation}
	\begin{aligned}
		\rho _{21D}=\frac{1}{\sqrt{\pi}u}\int_{-3u}^{3u}{\rho _{21}\left( \Delta _{p}^{'},\Delta _{c}^{'} \right) e^{-\frac{v^2}{u^2}}dv},
	\end{aligned}
\end{equation}
where $u =\sqrt{k_BT/m}$, $m$ is the atom's mass, $k_B$ is Boltzmann constant, and $T$ is the thermodynamic temperature. At room temperature, assume $T = 303.15 \ \text{K}$. Due to the atomic thermal motion, the probe and coupling light detuning $\Delta _{p}^{'}$ and $\Delta _{c}^{'}$ can be modified by the following

\begin{equation}
	\begin{aligned}
		\Delta _{p}^{'}=\Delta _p-\frac{2\pi}{\lambda _p}v,
		\\
		\Delta _{c}^{'}=\Delta _c+\frac{2\pi}{\lambda _c}v.
	\end{aligned}
\end{equation}

%\newpage
\normalem
\bibliographystyle{IEEEtran}
\bibliography{myref2}

\end{document}